\documentclass{aa}  
\usepackage{natbib}

\usepackage{color}

\usepackage{graphicx}
\usepackage{txfonts}
\begin{document}
\title{Origin of Black Hole Spin in Lower-Mass-Gap Black Hole-Neutron Star Binaries}

\author{Ying Qin \inst{1}
          \and
Zhen-Han-Tao Wang \inst{2}
          \and
Georges Meynet \inst{3,4}
          \and
Rui-Chong Hu \inst{5,6}
           \and
Chengjie Fu \inst{1}
           \and
Xin-Wen Shu \inst{1}
           \and
Zi-Yuan Wang \inst{1}
            \and
Shuang-Xi Yi\inst{7}  
          \and
Qing-Wen Tang\inst{8}
          \and
Han-Feng Song \inst{9}
           \and
En-Wei Liang \inst{2}           
}

\institute{
            Department of Physics, Anhui Normal University, Wuhu, Anhui, 241002, China\\
              \email{yingqin2013@hotmail.com}   
         \and    
Guangxi Key Laboratory for Relativistic Astrophysics, School of Physical Science and Technology, Guangxi University, Nanning 530004, China
         \and
         Département d’Astronomie, Université de Genève, Chemin Pegasi 51, 1290 Versoix, Switzerland
        \and 
        Gravitational Wave Science Center (GWSC), Université de Genève, 24 quai E. Ansermet, 1211 Geneva, Switzerland
        \and
        Nevada Center for Astrophysics, University of Nevada, Las Vegas, NV 89154, USA
        \and 
        Department of Physics and Astronomy, University of Nevada, Las Vegas, NV 89154, USA
        \and 
        School of Physics and Physical Engineering, Qufu Normal University, Qufu, Shandong 273165, China
        \and
         Department of Physics, School of Physics and Materials Science, Nanchang University, Nanchang 330031, China
        \and
         College of Physics, Guizhou University, Guiyang, Guizhou 550025, PR China
        }

\abstract
{During the fourth observing run, the LIGO-Virgo-KAGRA Collaboration reported the detection of a coalescing compact binary (GW230529$_{-}$181500) with component masses estimated at $2.5-4.5\, M_\odot$ and $1.2-2.0\, M_\odot$ with 90\% credibility. Given the current constraints on the maximum neutron star (NS) mass, this event is most likely a lower-mass-gap (LMG) black hole-neutron star (BHNS) binary. The spin magnitude of the BH, especially when aligned with the orbital angular momentum, is critical in determining whether the NS is tidally disrupted. An LMG BHNS merger with a rapidly spinning BH is an ideal candidate for producing electromagnetic counterparts. However, no such signals have been detected. In this study, we employ a detailed binary evolution model, incorporating new dynamical tide implementations, to explore the origin of BH spin in an LMG BHNS binary. If the NS forms first, the BH progenitor (He-rich star) must begin in orbit shorter than 0.35 days to spin up efficiently, potentially achieving a spin magnitude of $\chi_{\rm BH} > 0.3$. Alternatively, if a non-spinning BH (e.g., $M_{\rm BH} = 3.6\, M_\odot$) forms first, it can accrete up to $\sim 0.2\, M_\odot$ via Case BA mass transfer (MT), reaching a spin magnitude of $\chi_{\rm BH} \sim 0.18$ under Eddington-limited accretion. With a higher Eddington accretion limit (i.e., 10.0 $\Dot{M}_{\rm Edd}$), the BH can attain a significantly higher spin magnitude of $\chi_{\rm BH} \sim\,0.65$ by accreting approximately $1.0\, M_\odot$ during Case BA MT phase.}


\maketitle
\section{Introduction}\label{sect1}
Mergers of black hole--neutron star (BHNS) systems are prime targets for gravitational-wave (GW) detection by Advanced LIGO \citep{aasi2015}, Advanced Virgo \citep{acernese2015}, and KAGRA \citep{Somiya2012,aso2013} observatories. The LIGO-Virgo-KAGRA (LVK) Collaboration first detected two high-confidence GWs from BHNS mergers, GW200105 and GW200115 \citep{Abbott2021BHNS,nitz2021}, at the end of the third observing run (O3). On May 29, 2023, during the fourth observing run (O4), GW230529$_{-}$181500 (abbreviated as GW230529) was detected at 18:15:00 UTC with high significance by the LIGO Livingston observatory, while the other observatories were offline or lacked the sensitivity to detect this signal \citep{GW230529}. The LVK reported the two component masses of GW230529 to be $m_1 = 3.6_{-1.2}^{+0.8} M_\odot$ and $m_2 = 1.4_{-0.2}^{+0.6} M_\odot$ (90\% confidence intervals) by adopting a high-spin secondary prior.

Dynamical mass measurements of low-mass BH X-ray binary population have constrained the lower edge of stellar-mass BHs to be around 5 $M_\odot$ \citep{Bailyn1998,Ozel2010,Farr2011}. On the other hand, the maximum mass for nonrotating NSs has been inferred to be $2-3\, M_\odot$ \citep{Rhoades1974,Kalogera1996,Muller1996,Ozel2016,Margalit2017,Ai2020,Shao2020,Fan2024}. Combining the heaviest NSs and the lightest BHs suggests an absence of BHs within the mass range of $2.5-5\, M_\odot$, commonly referred to as the lower mass gap \cite[LMG, see a recent review paper from][]{shao2022RAA}. Therefore, considering current constraints on the maximum NS, the most probable interpretation of GW230529 is an LMG BHNS merger. \footnote{It is worth noting that \cite{Wangsong2024NA} recently reported a potential mass-gap BH in a wide binary with a circular orbit.}  For BHNS mergers, the effective inspiral spin $\chi_{\rm eff}$ that can be directly constrained by the GW signal, is defined as 

\begin{equation}
    \chi_{\rm eff} = \frac{m_{\rm BH} \boldsymbol{\chi}_{\rm NS} + m_{\rm NS} \boldsymbol{\chi}_{\rm BH}}{m_{\rm BH} + m_{\rm NS}} \cdot \hat{L}_N,
\end{equation}
where $m_{\rm BH}$ and $m_{\rm NS}$ are the BH and NS masses, $\boldsymbol{\chi}_{\rm BH}$ and $\boldsymbol{\chi}_{\rm NS}$ are their corresponding dimensionless spins, and $\hat{L}_N$ is the unit vector of the orbital angular momentum. Given the measurement of $\chi_{\rm eff} = -0.10_{-0.17}^{+0.12}$ for GW230529, its primary 
was inferred to have the spin component $\chi_{\rm 1,z} = -0.1_{-0.35}^{+0.19}$, which is typically low.

Instead of directly plunging into the BH, the NS in a BHNS merger can, under certain circumstances, be tidally disrupted by the BH's strong gravitational field. This tidal disruption generates remnant baryonic material outside the BH, potentially powering a range of electromagnetic (EM) counterparts, including a short gamma-ray burst \citep[e.g.,][]{Paczynski1991,Narayan1992,Zhang2018} and kilonova emission \citep[e.g.,][]{Li1998,Metzger2010,Zhu2020}. The intrinsic brightness of the EM emission in BHNS mergers is significantly influenced by the amount of mass ejected during the tidal disruption process. This mass ejection is determined by the system's properties, including the BH mass, the NS mass, the NS Equation-of-State (EoS), and the magnitude and orientation of the spins of both objects \citep[e.g.,][]{Kyutoku2015,Foucart2018,Zhu2021a,Zhu2021b}. Recent investigations suggested a probability of approximately 2\% to 41\% for kilonova emission from GW230529, depending on the EoS of the NS \citep{Kunnumkai2024}. In particular, an NS is more likely to be tidally disrupted by a low-mass BH with a high spin \citep{Hu2022}. Consequently, an LMG BHNS binary with a high BH spin is a promising candidate for producing significant tidal disruption. 

Binary population synthesis models using ``\texttt{delayed}'' supernova (SN) prescription \citep{Fryer2012} can produce BHNS binaries with the primary BH in the LMG \citep{Belczynski2012,Zevin2020,Broekgaarden2021,Chattopadhyay2021,Olejak2022,Zhu2024GW230529}. If the first-born compact object is an NS, it may accrete enough mass from its companion to trigger an accretion-induced collapse, potentially forming BHNS mergers like the source of GW230529 \citep{Zhu2024SuperEdd}. In the context of the field binary evolution, a BH can obtain a high spin either through tidal spin-up on its progenitor \cite[e.g.,][]{Detmers2008,Qin2018,Qin2019,Bavera2020,Hu2022,Hu2023,Qin2023}, or super-Eddington accretion from its companion star \cite[e.g.,][]{Bavera2021,Qin2022,Shao2022,Wang2024}. 

In this \textit{Letter}, we aim to study the origin of BH spin in a LMG BH and NS binary. To this end, we first study an NS + He-rich star interacting through tides, as well as a BH + He-rich star interacting through tides and mass transfer. Second, we identify those models that may be reasonable scenarios for the system GW230529. The structure of this paper is as follows. We briefly describe in Section \ref{sect2} the methodology used in our study. Subsequently in Section \ref{sect3}, we present the main results of our findings. Finally, the conclusions with some discussion are summarized in Section \ref{sect4}.

\section{Methods}\label{sect2}
We adopt the released version r15140 of the \texttt{MESA} stellar evolution code \citep{Paxton2011,Paxton2013,Paxton2015,Paxton2018,Paxton2019,Paxton2023} to conduct the detailed binary modeling. Following the methodology of recent studies \citep[e.g.,][]{Tassos2023,lv2023,Wang2024,Zhu2024GW230529}, we create single He-rich stars at the zero-age helium main sequence and then relax them to reach the thermal equilibrium (defined as when the ratio of He-burning luminosity to total luminosity exceeds 99\%). Convection is modeled using the mixing-length theory \citep{MLT1958}, with a mixing-length of $\alpha_{\rm mlt}=1.93$, while semiconvection is treated with an efficiency parameter of $\alpha_{\sc}=1.0$ \citep{Langer1983}. The Ledoux criterion is adopted to determine convective zone boundaries, and step overshooting is modeled as an extension with $\alpha_{\rm p} = 0.1 H_{\rm p}$, where $H_{\rm p}$ represents the pressure scale height at the Ledoux boundary. For nucleosynthesis calculations, we utilize the \texttt{approx21.net} network.

Rotational mixing and angular momentum transport are treated as diffusive processes \citep{Heger2000}, incorporating the effects of the Goldreich–Schubert–Fricke instability, Eddington–Sweet circulations, as well as both secular and dynamical shear mixing. We apply diffusive element mixing from these processes using an efficiency parameter of $f_{\rm c}=1/30$ \citep{Chaboyer1992,Heger2000}. Given the sensitivity of the $\mu$-gradient to rotationally induced mixing, we mitigate it by applying a reduction factor of $f_\mu = 0.05$, as recommended in \cite{Heger2000}. In our simulations, we evolve He-rich stars with initial metallicity of $Z=Z_{\odot}$, where $Z_{\odot}$ is set to 0.0142 \citep{Asplund2009}.

We model the wind mass-loss of He-rich stars following \cite{Hu2022} and account for the rotationally-enhanced
mass-loss using the formulation from \citep{Heger1998,Langer1998}:
    \begin{equation}\label{ml}
    \centering
    \dot{M}(\omega)= \dot{M}(0)\left(\frac{1}{1-\omega/\omega_{\rm crit}}\right)^\xi,
    \end{equation}
where $\omega$ and $\omega_{\rm crit}$ are the angular and critical angular velocities at the stellar surface, respectively. The critical angular velocity is defined as $\omega_{\rm crit}^2 = (1- L/L_{\rm Edd})GM/R^3$, with $L$, $M$, $R$ representing the star's total luminosity, mass, and radius, respectively; $L_{\rm Edd}$ is the Eddington luminosity, and $G$ the gravitational constant. The exponent $\xi = 0.43$ is adopted from \citet{Langer1998}. It is important to note that we do not account for gravity-darkening effects, as discussed in \cite{Maeder2000}.

Stars have extended atmospheres, allowing mass transfer (MT) to occur through the first Lagrangian point ($L_1$) even when the star remains within its Roche lobe \citep[i.e., $R_1 < R_{\rm RL}$, as described by the Ritter scheme,][]{Ritter1988}. \cite{Kolb1990} further extended the Ritter scheme to cases where $R_1 > R_{\rm RL}$. In this work, we model MT using the Kolb scheme \citep{Kolb1990} and adopt the implicit MT method \citep{Paxton2015}. For simplicity, we assume conservative mass transfer through $L_1$ between a He-rich star and its companion. 

For tidal interactions, we apply dynamical tides to He-rich stars with radiative envelopes \citep{Zahn1977}. The synchronization timescale is calculated using the prescriptions from \cite{Zahn1977,Hut1981,Hurley2002}, and the tidal torque coefficient $E_2$ is adopted from the updated fitting formula in \cite{Qin2018}. Recent studies by \cite{Sciarini2024} have highlighted inconsistencies in implementing dynamic tides compared to \cite{Zahn1977}. In this work, we use the correction proposed by \cite{Sciarini2024} and 
provide detailed explanations in the Appendix for readers of interest. 

Given the negligible natal spin of the first-born BH in binaries \cite[e.g.,][]{Fragos2015,Qin2018,Fuller2019,Belczynski2020}, we assume that the BH is initially non-rotating. As the BH accretes material through a disk at a rate $\dot{M}_{\rm acc}$, its mass increases according to the following growth rate:

\begin{equation}
\label{equ:BHMass}
\dot{M}_{\rm BH} = (1- \eta) \dot{M}_{\rm acc},
\end{equation}
where $\eta$ is the radiation efficiency determined by the innermost stable circular orbit. As the BH's mass increases, its spin evolves according to \citep{Bardeen1970,Bardeen1972,Podsiadlowski2003,Marchant2017}:  

\begin{equation} \label{equ:BHSpin}
    \chi_{\rm BH} = \sqrt{\frac{2}{3}} \frac{M_{\rm BH,init}}{M_{\rm BH}}\left( 4 -\sqrt{18\left ( \frac{M_{\rm BH,init}}{M_{\rm BH}} \right )^2 - 2} \right),  
\end{equation}
for $M_{\rm BH} < \sqrt{6} M_{\rm BH,init}$, where $\eta = 1 - \sqrt{1 - \left ( {M_{\rm BH}}/{3M_{\rm BH,init}} \right )^2}$. 

The standard Eddington accretion rate, which represents the critical rate at which the outward force from radiation pressure balances the inward gravitational pull, is defined as:
\begin{equation}\label{m_edd}
\dot{M}_{\rm Edd} =  \frac{4\pi G M_{\rm BH}}{\kappa c \eta},
\end{equation}
where $c$ is the speed of light, $G$ is the gravitational constant, and $\kappa$ is the opacity, primarily determined by pure electron scattering. The opacity is given by $\kappa = 0.2(1+X),{\rm cm}^2,{\rm g}^{-1}$, where $X$ is the hydrogen mass fraction \citep[e.g.,][]{Kippenhahn1990}.

We evolve He-rich stars until central carbon exhaustion is achieved. To calculate the baryonic remnant mass ($M_{\rm bar}$), we use the ``\texttt{delayed}'' SN prescription \citep{Fryer2012}. The gravitational mass ($M_{\rm rem}$) for neutron stars (NSs) is then determined using the following relation \citep{Lattimer1989,Timmes1996}:
\begin{equation}\label{ml}
\centering
     M_{\rm bar} - M_{\rm rem} = 0.075\, M_{\rm rem}^2,
\end{equation}
while for BHs we follow the method as in \cite{Fryer2012} to approximate the $M_{\rm rem}$ using
\begin{equation}\label{ml}
\centering
     M_{\rm rem} = 0.9\, M_{\rm bar}.
\end{equation}
Additionally, we account for the neutrino loss mechanism described by \cite{Zevin2020}. In this study, we assume a maximum NS mass of 2.5,$M_{\odot}$.

\section{Results}\label{sect3}
\subsection{Tidal spin-up}
\begin{figure}[h]
     \centering
     \includegraphics[width=\columnwidth]{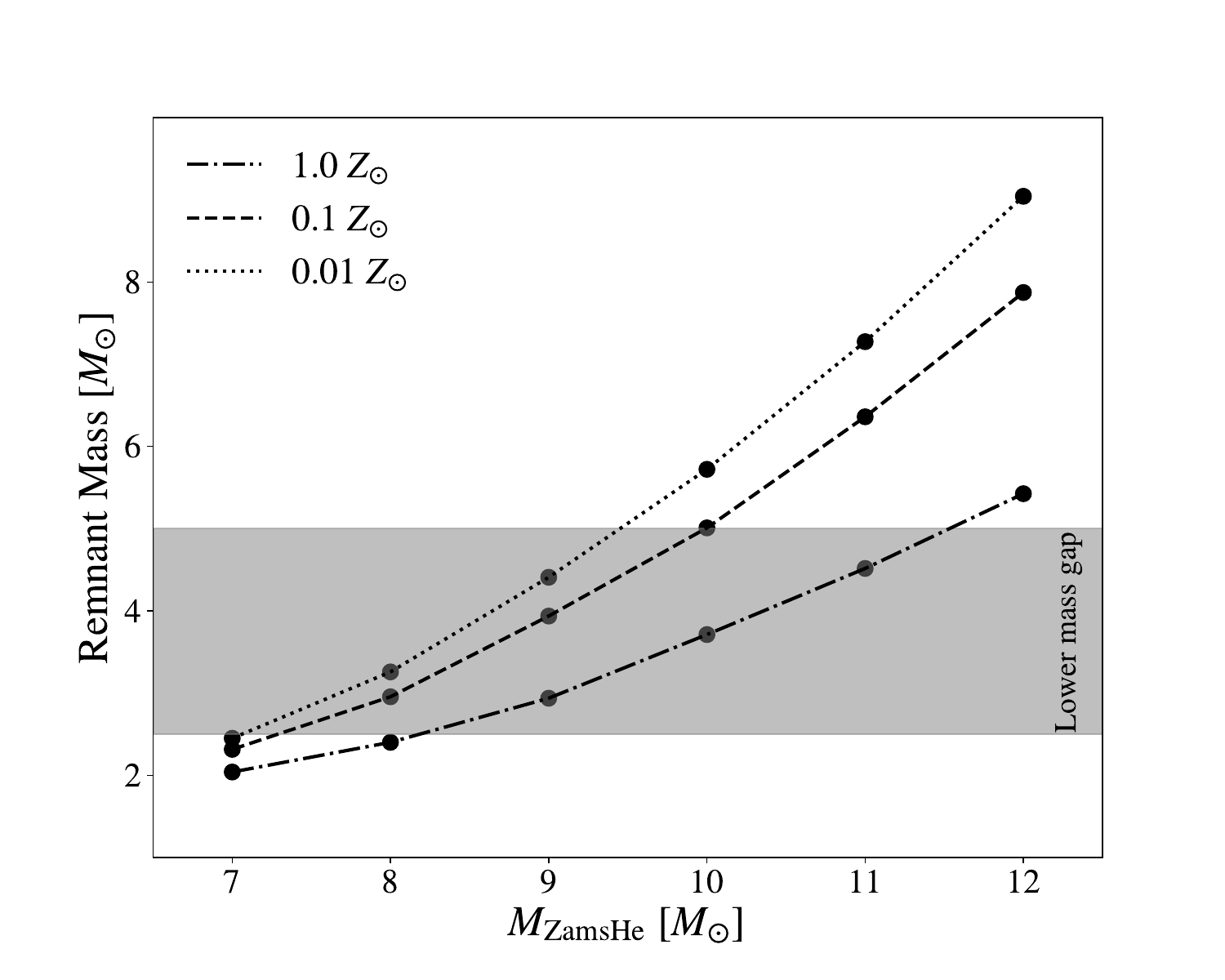}
     \caption{Remnant mass as a function of the initial mass of non-rotating He-rich stars at three distinct metallicities, represented by different line styles: dash-dotted line for 1.0 $Z_\odot$, dashed line for 0.1 $Z_\odot$, and dotted line for 0.01 $Z_\odot$. The LMG is indicated by the grey-shaded region.}
     \label{fig1}
\end{figure}

In the scenario of tidal spin-up, we are focused on systems composed of an NS and a He-rich star in a close binary. As investigated previously by \cite{Detmers2008,Qin2018,Hu2022}, a He-rich star in a close binary system can be efficiently spun up by tidal interactions with its companion. This process can potentially lead to the formation of a fast-spinning NS or BH, with the outcome primarily dependent on the initial mass of the He-rich star and its metallicity \citep[see Figure 3 in][]{lv2023}. Figure \ref{fig1} illustrates the remnant mass as a function of the initial He-rich star mass across different metallicities (1.0 $Z_\odot$, 0.1 $Z_\odot$, and 0.01 $Z_\odot$). At lower metallicities (0.1 $Z_\odot$ and 0.01 $Z_\odot$), He-rich stars are expected to retain more mass due to the reduced strength of metallicity-dependent winds \citep{vink2001}. For BHs formed at solar metallicity within the LMG ($2.5 - 5.0\, M_\odot$), the initial mass of He-rich stars needs to 
range from approximately 8.0 $M_\odot$ to $\sim\, 11.5\, M_\odot$ (from $\sim\, 7.0\, M_\odot$ to $\sim\, 10.0\, M_\odot$ at 0.1 $Z_\odot$ and from $\sim\, 7.0\, M_\odot$ to $\sim\, 9.5\, M_\odot$ at 0.01 $Z_\odot$). 

We carry out detailed binary modeling of He-rich stars to track the evolution of their angular momentum, ultimately determining the spin of the resulting BHs. The initial mass range of He-rich stars is selected to ensure the formation of BHs within the LMG. We model NS as a point mass with $M_{\rm NS} = 1.4\, M_\odot$ and evolve He-rich star with initial masses ($M_{\rm ZamHe}$) linearly ranging from 7.0\,$M_\odot$ to 12.0\,$M_\odot$ at a step of 1.0\,$M_\odot$. The initial orbital period spans from approximately 0.06 days (the point where He-rich stars begin to overflow their Roche lobes) to 2.2 days \citep[beyond which tidal interactions become negligible; see detailed testing in][]{Qin2018}, with a logarithmic spacing of $\Delta\log(P/{\rm days}) \approx 0.16\,{\rm dex}$. We evolve He-rich stars with different initial metallicities (1.0 $Z_\odot$, 0.1 $Z_\odot$, and 0.01 $Z_\odot$) until central carbon depletion. In all binary modelings, the initial angular velocities of He-rich stars are synchronized with the orbital motion. 

\begin{figure*}[h]
     \centering
     \includegraphics[width=0.99\textwidth]{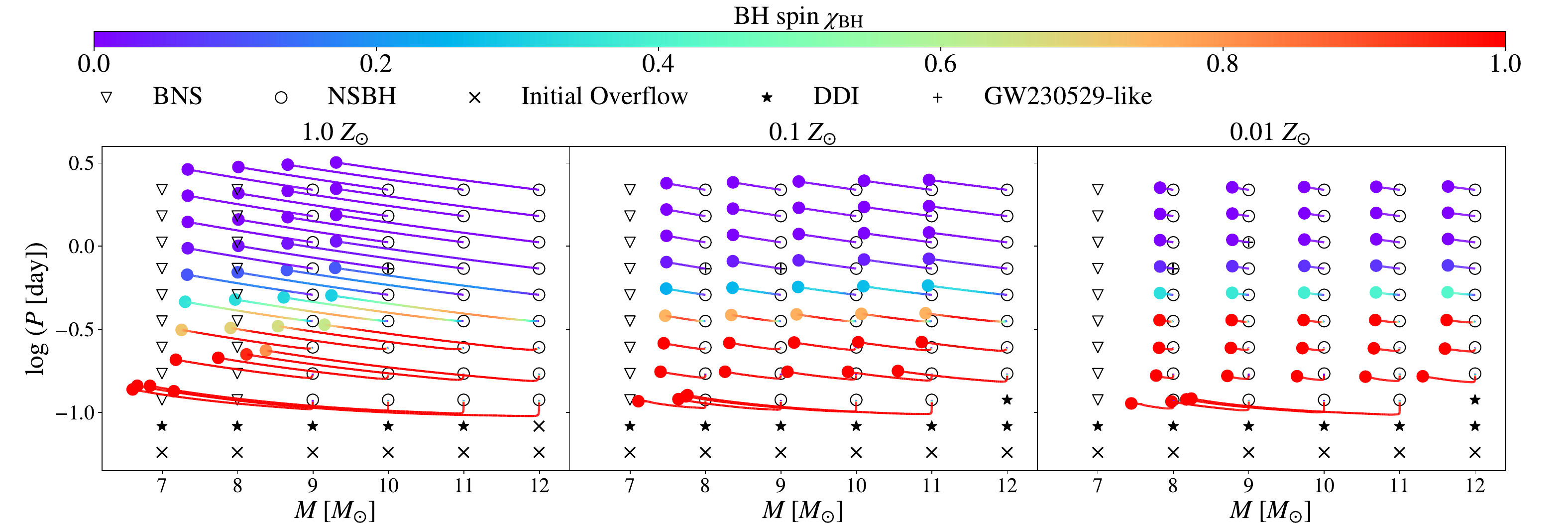}
     \caption{BH dimensionless spin $\chi_{\rm BH}$ as a function of the He-rich star mass and the orbital period. Black symbols represent the initial parameters (He-rich star mass and orbital period), while colored symbols indicate the final parameters. The color gradient along the line traces the evolution of $\chi_{\rm BH}$, assuming that the He-rich star collapses directly into a BH without losing mass or angular momentum. Triangles represent BNS systems, circles denote NSBH systems, stars indicate dynamical delayed instability (DDI), and crosses mark initial overflow cases. The three columns correspond to different metallicities: \textit{left:} 1.0 $Z_\odot$; \textit{middle:} 0.1 $Z_\odot$; \textit{right:} 0.01 $Z_\odot$. The plus symbol refers to the systems that might be representative of the progenitors of GW230529.} 
     \label{fig2}
\end{figure*}

Figure \ref{fig2} displays the spin of BHs formed from the direct collapse of He-rich stars under various conditions. The left panel shows the results for He-rich stars with an initial metallicity of 1.0 $Z_\odot$. For systems in wider orbits ($\log(P/{\rm days}) \gtrsim -0.25$), tidal forces are too weak to significantly spin up He-rich stars, resulting in BHs with low spin ($\chi_{\rm BH} < 0.2$). To achieve a moderate BH spin (i.e., $\chi_{\rm BH} > 0.3$), the initial orbital period must be shorter than approximately 0.35 days (log$P \sim -0.46$). Stellar winds from He-rich stars typically reduce their mass, leading to a widening of the binary system. However, He-rich stars with lower initial metallicities can form double compact objects in closer orbits due to weaker winds. These low-metallicity stars (see the middle and right panels) can retain more angular momentum, forming fast-spinning BHs if they are sufficiently massive. Additionally, NSBH systems (where the NS forms first, as opposed to BHNS systems where the BH forms first) formed at low metallicities are expected to merge more quickly due to gravitational wave emission \citep{Peters1964}. We note that the parameter space (see the plus symbol) that leads to the formation of GW230529 is very limited when considering its low BH spin.

\subsection{Accretion-induced spin-up}
\begin{figure}[h]
     \centering
     \includegraphics[width=\columnwidth]{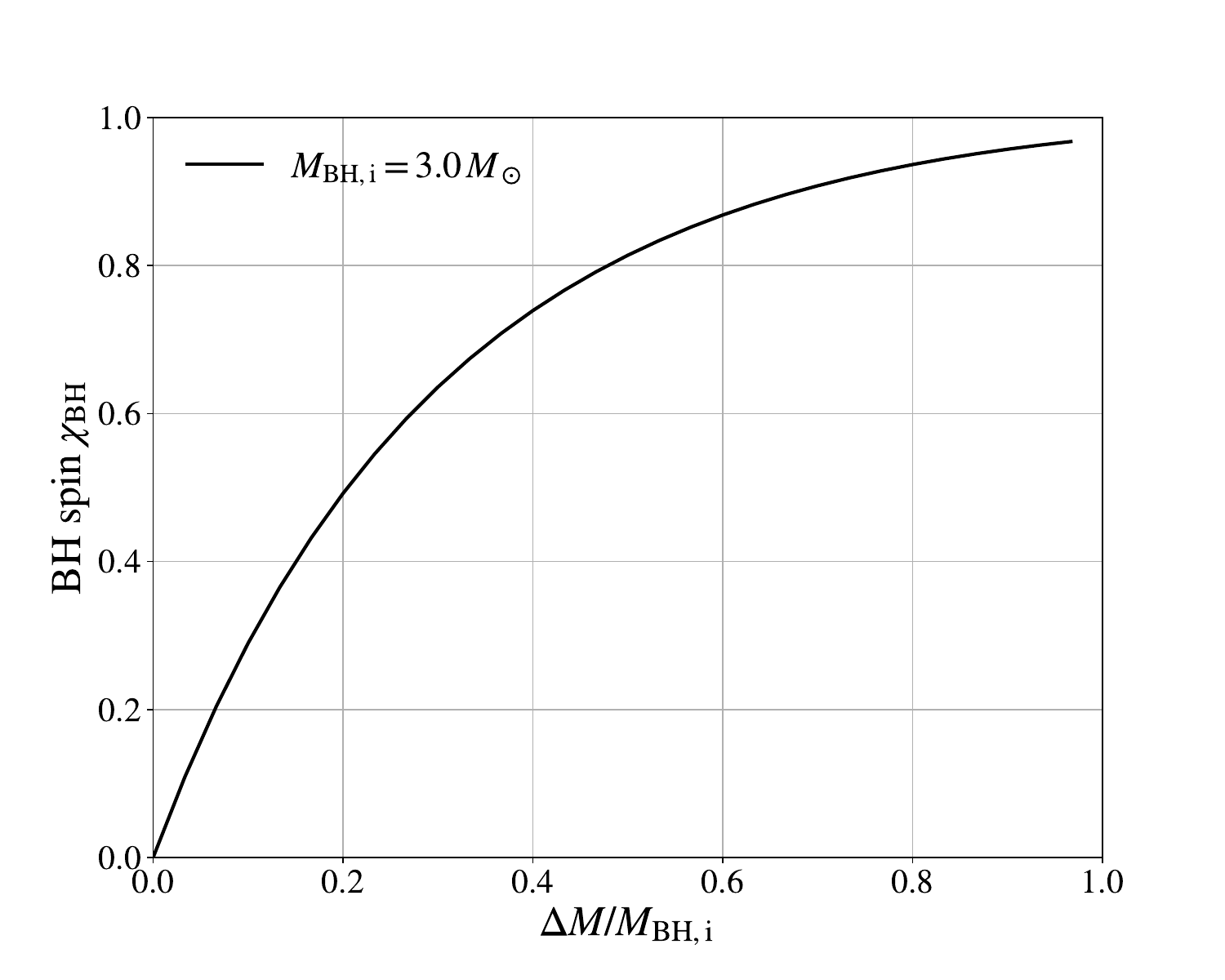}
     \caption{BH dimensionless spin $\chi_{\rm BH}$ as a function of the ratio of accreted mass to its initial mass.}
     \label{fig3}
\end{figure}

 In the scenario of accretion-induced spin-up, we consider a close binary system consisting of a BH and a He-rich star. As an example, Figure \ref{fig3} illustrates how the spin of the BH increases as it accretes material. Previous studies have shown that super-Eddington accretion can well explain the high spin of BH in high-mass X-ray binaries \citep{Qin2022} and in double BH systems \citep[e.g.,][]{Bavera2021,Shao2022,Zevin2022,Briel2023}. Observational evidence for super-Eddington accretion has been identified in several BH X-ray binaries, including examples such as V404 Cygni \citep{Motta2017} and GRO J1655-40 \citep{Neilsen2016}. In our binary modeling, we allow the accretion rate to exceed the standard Eddington limit by up to 10 times, i.e., 10 $\dot{M}_{\rm Edd}$.

\begin{figure}[h]
     \centering
     \includegraphics[width=\columnwidth]{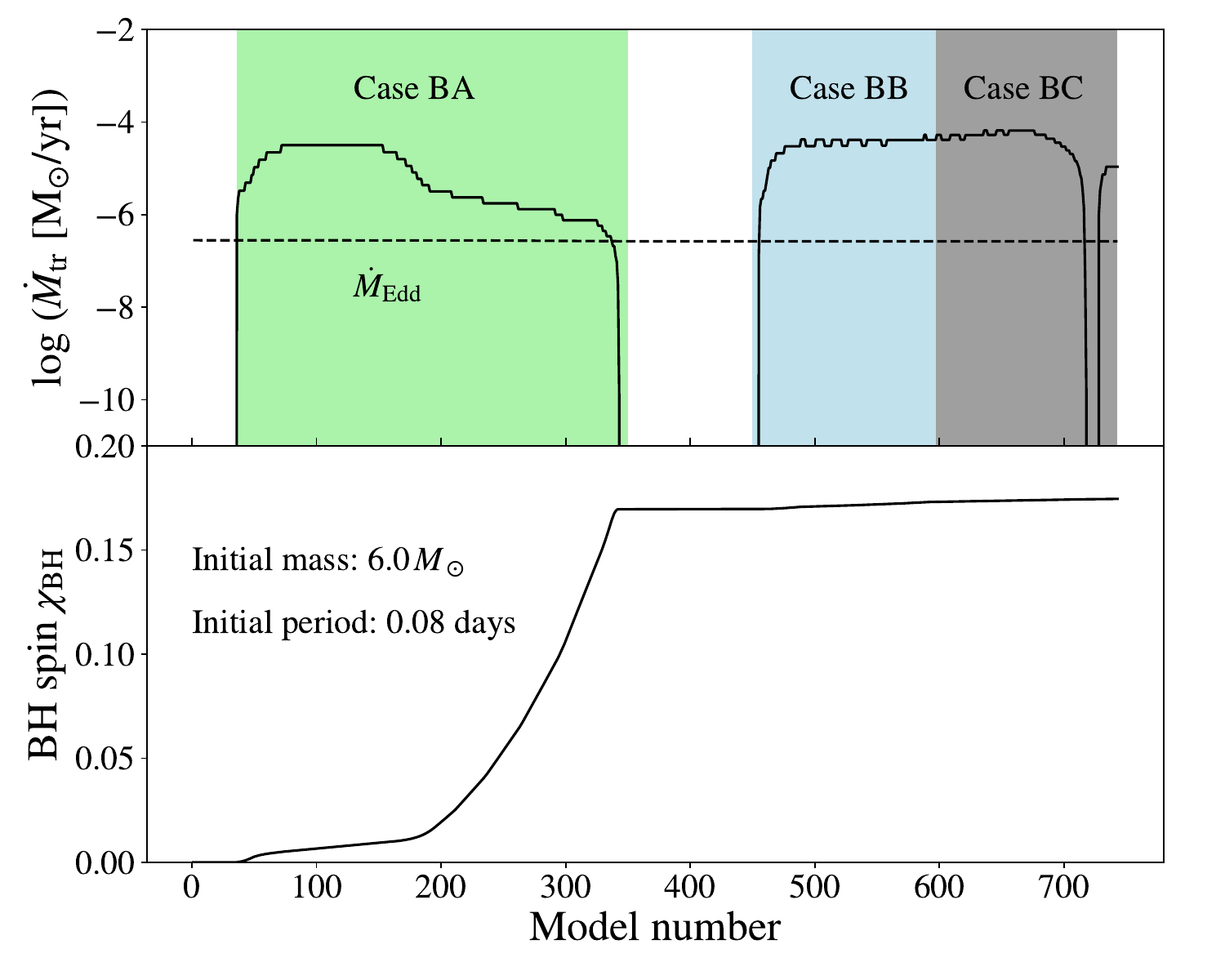}
     \caption{MT rate (upper panel) and BH spin $\chi_{\rm BH}$ (lower panel) as functions of the model number. The dashed line indicates the standard Eddington accretion rate (1.0 $\Dot{M}_{\rm Edd}$). In the binary sequence shown, the initial masses of the BH and He-rich star are $3.6\, M_\odot$ and $6.0\, M_\odot$, respectively, with an initial orbital period of 0.08 days. Different MT phases are highlighted in various colors: Case BA (green), Case BB (blue), and Case BC (grey).}
     \label{fig4}
\end{figure} 

Before delving into the results of the grid study, we first present our findings on the evolution of a specific binary sequence, as shown in Figure \ref{fig4}. This close binary system consists of a 3.6 $M_\odot$ BH and a 6.0 $M_\odot$ He-rich star, with an initial orbital period of 0.08 days. The upper panel illustrates that the system undergoes Case BA, BB, and BC MT phases. During the Case BA MT phase, the MT rate initially rises above $10^{-5} M_\odot/$yr before gradually decreasing to the Eddington-limited accretion rate. Under the assumption of standard Eddington accretion, we consider any excess material to be lost from the vicinity of the accretor as an isotropic wind, carrying the specific orbital angular momentum of the accretor. Notably, the BH spin increases to $\chi_{\rm BH} \sim 0.17$. The system then becomes detached as the He-rich star temporarily contracts at the end of the core-helium burning phase. The He-rich star subsequently expands, initiating the Case BB and Case BC MT phases, during which the BH spin increases more slowly due to the limited duration of the accretion period.

As illustrated in Figure \ref{fig1}, we focus on He-rich stars with solar metallicity and masses ranging from 3.0 to 8.0 $M_\odot$, where NSs are expected to form. The BH is modeled as a point mass with $M_{\rm BH} = 3.6\, M_\odot$. The upper limit for the initial orbital period is set to 10 days, below which no MT occurs through the $L_1$ point \citep[see recent studies on similar MT phases in][]{Zhang2023}. Conversely, the lower limit corresponds to systems where Roche-lobe overflow begins in the initial model. In the left panel of Figure \ref{fig5}, we show how the BH mass evolves as a function of the He-rich star's mass and the orbital period. Assuming Eddington-limited accretion, the BH can accrete up to approximately 0.2 $M_\odot$ during the Case BA MT phase, reaching a maximum mass of $M_{\rm BH} \approx 3.8\, M_\odot$. This accretion adds specific angular momentum, spinning up the BH to a dimensionless spin parameter $\chi_{\rm BH} \approx 0.18$. If the BH accretes material beyond the standard Eddington limit, it can significantly increase both its mass and spin (also see Figure 3). The right panels of Figures \ref{fig5} and \ref{fig6} display the evolution of BH mass and spin within the given parameter space, allowing for up to 10 $\dot{M}_{\rm Edd}$ in our binary modeling. Among all the binary sequences, the BH can increase its mass by up to approximately 1.0 $M_\odot$ through the Case BA MT phase, resulting in an efficient spin-up to $\chi_{\rm BH} \approx 0.65$. In both cases, the parameter space (see the plus symbol) that potentially produces GW230529 is much larger compared with the scenario of tidal spin-up.


\section{Conclusions and discussion}\label{sect4}
In this study, we use detailed binary evolution tools to investigate the origins of BH spin in lower-mass-gap (LMG) BH and NS binary systems. We explore two primary formation scenarios: tidal spin-up of the BH progenitor and accretion-induced spin-up. To address this, we conduct a parameter space study across two grids of He-rich stars in binary systems, analyzing various conditions.

First, we evolve single He-rich stars to determine the mass range in which BHs are expected to form within the LMG across three different metallicities. With the new implementation of the dynamical tides as suggested in \cite{Sciarini2024}, we find that He-rich stars must be in closer orbits to achieve significant spin-up. In the tidal spin-up scenario, we perform a grid of detailed binary models involving a He-rich star and an NS in close orbits. At solar metallicity, in systems with wider orbits ($\log(P/{\rm days}) \gtrsim -0.25$), tidal forces are insufficient to spin up He-rich stars, resulting in low-spin BHs ($\chi_{\rm BH} < 0.2$). To achieve moderate BH spin ($\chi_{\rm BH} > 0.3$), the initial orbital period needs to be shorter than approximately 0.35 days ($\log(P/{\rm days}) \sim -0.46$). At lower metallicities, weaker mass-loss winds cause the systems to become more compact, resulting in the formation of faster BH binaries. In this scenario, it is noted that the spin of the BH formed in the LMG can span the entire range from zero to maximally spinning. Recent findings by \cite{Ma2023} show that tidal interactions are predominantly driven by standing gravity modes, as opposed to traveling waves, which were previously suggested by \cite{Zahn1977}. As a result, earlier estimates of tidal spin-up may have been overestimated.

In the accretion-induced spin-up scenario, we explore a grid of models involving a BH with $M_{\rm BH} = 3.6\, M_\odot$ and He-rich stars within the mass range of 3.0 to 8.0 $M_\odot$. Assuming Eddington-limited accretion, the BH can accrete up to approximately 0.2 $M_\odot$ during the Case BA MT phase, resulting in a spin magnitude of $\chi_{\rm BH} \sim 0.18$. However, under a higher accretion limit of 10.0 $\Dot{M}_{\rm Edd}$, the BH can achieve a significantly higher spin magnitude of $\chi_{\rm BH} \sim 0.65$ by accreting about 1.0 $M_\odot$ during the same MT phase. In this scenario, the spin of the BH formed in the LMG is typically small even with a moderate Eddington accretion limit when compared with the case of tidal spin-up. Only in a few systems can BHs attain a high spin magnitude. Notably, the spin of a BH can decrease due to the launching of a Blandford-Znajek jet \citep{Blandford1977} during periods of hyperaccretion \citep{Lei2017, Wang2024NA}. For merging binary BHs formed through isolated binary evolution channels, \cite{Monica2021} found that stable MT, rather than the classical common envelope channel, may play a dominant role. In contrast, they found that binary neutron star mergers are primarily formed through the common envelope channel, rather than stable MT \citep{Monica2023}.

The formation channels explored in this study could be relevant to the origin of GW230529 \citep{GW230529}. Using the population synthesis code \texttt{COMPAS} \cite[][]{Stevenson2017,Vigna2018,Neijssel2019,TeamCOMPAS2022}, \cite{Zhu2024GW230529} suggested that GW230529 likely originated from the isolated massive binary evolution channel. In support of this, \cite{Chandra2024} found that the first-born BH in GW230529 is consistent with the classical isolated binary evolution scenario. This aligns with earlier studies indicating that BHs typically form first \citep{Xing2024}, largely due to rotation-dependent accretion efficiency onto nondegenerate stars \citep{Demink2013}. Although the possibility of GW230529 forming through dynamical assembly cannot be entirely ruled out, current estimates suggest that the merger rate through this channel is particularly low \cite[e.g., $\lesssim 1\, \rm Gpc^{-3} yr^{-1}$, see][]{Ye2024}. As more data is gathered by the LVK Collaboration, additional BH and NS mergers involving BH within the LMG are likely to be reported. 

\begin{acknowledgements}
We thank the referee for constructive comments that helped improve the manuscript. We also thank Bing Zhang and Jin-Ping Zhu for their helpful comments. Y.Q. acknowledges support from the Anhui Provincial Natural Science Foundation (grant No. 2308085MA29) and the National Natural Science Foundation of China (grant No. 12473036). G.M. has received funding from the European Research Council (ERC) under the European Union’s Horizon 2020 research and innovation program (grant agreement No 833925, project STAREX). C.J.F is supported by the National Natural Science Foundation of China Grant No. 12305057. Q.W.T acknowledges support from Jiangxi Provincial Natural Science Foundation (grant Nos. 20242BAB26012 and 20224ACB211001). This work was partially supported by the National Natural Science Foundation of China (grant Nos. 12065017, 12192220, 12192221, 12133003, 12203101, U2038106, 12103003). All figures are made with the free Python module Matplotlib \citep{Hunter2007}.
\end{acknowledgements}
\bibliography{ref}

\begin{thebibliography}{104}
\expandafter\ifx\csname natexlab\endcsname\relax\def\natexlab#1{#1}\fi

\bibitem[{{Aasi} {et~al.}(2015){Aasi}, {Abbott}, {Abbott}, {Abbott}, {Abernathy}, {Ackley}, {Adams}, {Adams}, {Addesso}, {Adhikari}, {Adya}, {Affeldt}, {Aggarwal}, {Aguiar}, {Ain}, {Ajith}, {Alemic}, {Allen}, {Amariutei}, {Anderson}, {Anderson}, {Arai}, {Araya}, {Arceneaux}, {Areeda}, {Ashton}, {Ast}, {Aston}, {Aufmuth}, {Aulbert}, {Aylott}, {Babak}, {Baker}, {Ballmer}, {Barayoga}, {Barbet}, {Barclay}, {Barish}, {Barker}, {Barr}, {Barsotti}, {Bartlett}, {Barton}, {Bartos}, {Bassiri}, {Batch}, {Baune}, {Behnke}, {Bell}, {Bell}, {Benacquista}, {Bergman}, {Bergmann}, {Berry}, {Betzwieser}, {Bhagwat}, {Bhandare}, {Bilenko}, {Billingsley}, {Birch}, {Biscans}, {Biwer}, {Blackburn}, {Blackburn}, {Blair}, {Blair}, {Bock}, {Bodiya}, {Bojtos}, {Bond}, {Bork}, {Born}, {Bose}, {Brady}, {Braginsky}, {Brau}, {Bridges}, {Brinkmann}, {Brooks}, {Brown}, {Brown}, {Brown}, {Buchman}, {Buikema}, {Buonanno}, {Cadonati}, {Calder{\'o}n Bustillo}, {Camp}, {Cannon}, {Cao}, {Capano}, {Caride}, {Caudill}, {Cavagli{\`a}}, {Cepeda},
  {Chakraborty}, {Chalermsongsak}, {Chamberlin}, {Chao}, {Charlton}, {Chen}, {Cho}, {Cho}, {Chow}, {Christensen}, {Chu}, {Chung}, {Ciani}, {Clara}, {Clark}, {Collette}, {Cominsky}, {Constancio}, {Cook}, {Corbitt}, {Cornish}, {Corsi}, {Costa}, {Coughlin}, {Countryman}, {Couvares}, {Coward}, {Cowart}, {Coyne}, {Coyne}, {Craig}, {Creighton}, {Creighton}, {Cripe}, {Crowder}, {Cumming}, {Cunningham}, {Cutler}, {Dahl}, {Dal Canton}, {Damjanic}, {Danilishin}, {Danzmann}, {Dartez}, {Dave}, {Daveloza}, {Davies}, {Daw}, {DeBra}, {Del Pozzo}, {Denker}, {Dent}, {Dergachev}, {DeRosa}, {DeSalvo}, {Dhurandhar}, {D{\textasciiacute}{\i}az}, {Di Palma}, {Dojcinoski}, {Dominguez}, {Donovan}, {Dooley}, {Doravari}, {Douglas}, {Downes}, {Driggers}, {Du}, {Dwyer}, {Eberle}, {Edo}, {Edwards}, {Edwards}, {Effler}, {Eggenstein}, {Ehrens}, {Eichholz}, {Eikenberry}, {Essick}, {Etzel}, {Evans}, {Evans}, {Factourovich}, {Fairhurst}, {Fan}, {Fang}, {Farr}, {Farr}, {Favata}, {Fays}, {Fehrmann}, {Fejer}, {Feldbaum}, {Ferreira}, {Fisher},
  {Frei}, {Freise}, {Frey}, {Fricke}, {Fritschel}, {Frolov}, {Fuentes-Tapia}, {Fulda}, {Fyffe}, {Gair}, {Gaonkar}, {Gehrels}, {Gergely}, {Giaime}, {Giardina}, {Gleason}, {Goetz}, {Goetz}, {Gondan}, {Gonz{\'a}lez}, {Gordon}, {Gorodetsky}, {Gossan}, {Go{\ss}ler}, {Gr{\"a}f}, {Graff}, {Grant}, {Gras}, {Gray}, {Greenhalgh}, {Gretarsson}, {Grote}, {Grunewald}, {Guido}, {Guo}, {Gushwa}, {Gustafson}, {Gustafson}, {Hacker}, {Hall}, {Hammond}, {Hanke}, {Hanks}, {Hanna}, {Hannam}, {Hanson}, {Hardwick}, {Harry}, {Harry}, {Hart}, {Hartman}, {Haster}, {Haughian}, {Hee}, {Heintze}, {Heinzel}, {Hendry}, {Heng}, {Heptonstall}, {Heurs}, {Hewitson}, {Hild}, {Hoak}, {Hodge}, {Hollitt}, {Holt}, {Hopkins}, {Hosken}, {Hough}, {Houston}, {Howell}, {Hu}, {Huerta}, {Hughey}, {Husa}, {Huttner}, {Huynh}, {Huynh-Dinh}, {Idrisy}, {Indik}, {Ingram}, {Inta}, {Islas}, {Isler}, {Isogai}, {Iyer}, {Izumi}, {Jacobson}, {Jang}, {Jawahar}, {Ji}, {Jim{\'e}nez-Forteza}, {Johnson}, {Jones}, {Jones}, {Ju}, {Haris}, {Kalogera}, {Kandhasamy}, {Kang},
  {Kanner}, {Katsavounidis}, {Katzman}, {Kaufer}, {Kaufer}, {Kaur}, {Kawabe}, {Kawazoe}, {Keiser}, {Keitel}, {Kelley}, {Kells}, {Keppel}, {Key}, {Khalaidovski}, {Khalili}, {Khazanov}, {Kim}, {Kim}, {Kim}, {Kim}, {Kim}, {King}, {King}, {Kinzel}, {Kissel}, {Klimenko}, {Kline}, {Koehlenbeck}, {Kokeyama}, {Kondrashov}, {Korobko}, {Korth}, {Kozak}, {Kringel}, {Krishnan}, {Krueger}, {Kuehn}, {Kumar}, {Kumar}, {Kuo}, {Landry}, {Lantz}, {Larson}, {Lasky}, {Lazzarini}, {Lazzaro}, {Le}, {Leaci}, {Leavey}, {Lebigot}, {Lee}, {Lee}, {Lee}, {Leong}, {Levin}, {Levine}, {Lewis}, {Li}, {Libbrecht}, {Libson}, {Lin}, {Littenberg}, {Lockerbie}, {Lockett}, {Logue}, {Lombardi}, {Lormand}, {Lough}, {Lubinski}, {L{\"u}ck}, {Lundgren}, {Lynch}, {Ma}, {Macarthur}, {MacDonald}, {Machenschalk}, {MacInnis}, {Macleod}, {Maga{\~n}a-Sandoval}, {Magee}, {Mageswaran}, {Maglione}, {Mailand}, {Mandel}, {Mandic}, {Mangano}, {Mansell}, {M{\'a}rka}, {M{\'a}rka}, {Markosyan}, {Maros}, {Martin}, {Martin}, {Martynov}, {Marx}, {Mason}, {Massinger},
  {Matichard}, {Matone}, {Mavalvala}, {Mazumder}, {Mazzolo}, {McCarthy}, {McClelland}, {McCormick}, {McGuire}, {McIntyre}, {McIver}, {McLin}, {McWilliams}, {Meadors}, {Meinders}, {Melatos}, {Mendell}, {Mercer}, {Meshkov}, {Messenger}, {Meyers}, {Miao}, {Middleton}, {Mikhailov}, {Miller}, {Miller}, {Millhouse}, {Ming}, {Mirshekari}, {Mishra}, {Mitra}, {Mitrofanov}, {Mitselmakher}, {Mittleman}, {Moe}, {Mohanty}, {Mohapatra}, {Moore}, {Moraru}, {Moreno}, {Morriss}, {Mossavi}, {Mow-Lowry}, {Mueller}, {Mueller}, {Mukherjee}, {Mullavey}, {Munch}, {Murphy}, {Murray}, {Mytidis}, {Nash}, {Nayak}, {Necula}, {Nedkova}, {Newton}, {Nguyen}, {Nielsen}, {Nissanke}, {Nitz}, {Nolting}, {Normandin}, {Nuttall}, {Ochsner}, {O'Dell}, {Oelker}, {Ogin}, {Oh}, {Oh}, {Ohme}, {Oppermann}, {Oram}, {O'Reilly}, {Ortega}, {O'Shaughnessy}, {Osthelder}, {Ott}, {Ottaway}, {Ottens}, {Overmier}, {Owen}, {Padilla}, {Pai}, {Pai}, {Palashov}, {Pal-Singh}, {Pan}, {Pankow}, {Pannarale}, {Pant}, {Papa}, {Paris}, {Patrick}, {Pedraza}, {Pekowsky},
  {Pele}, {Penn}, {Perreca}, {Phelps}, {Pierro}, {Pinto}, {Pitkin}, {Poeld}, {Post}, {Poteomkin}, {Powell}, {Prasad}, {Predoi}, {Premachandra}, {Prestegard}, {Price}, {Principe}, {Privitera}, {Prix}, {Prokhorov}, {Puncken}, {P{\"u}rrer}, {Qin}, {Quetschke}, {Quintero}, {Quiroga}, {Quitzow-James}, {Raab}, {Rabeling}, {Radkins}, {Raffai}, {Raja}, {Rajalakshmi}, {Rakhmanov}, {Ramirez}, {Raymond}, {Reed}, {Reid}, {Reitze}, {Reula}, {Riles}, {Robertson}, {Robie}, {Rollins}, {Roma}, {Romano}, {Romanov}, {Romie}, {Rowan}, {R{\"u}diger}, {Ryan}, {Sachdev}, {Sadecki}, {Sadeghian}, {Saleem}, {Salemi}, {Sammut}, {Sandberg}, {Sanders}, {Sannibale}, {Santiago-Prieto}, {Sathyaprakash}, {Saulson}, {Savage}, {Sawadsky}, {Scheuer}, {Schilling}, {Schmidt}, {Schnabel}, {Schofield}, {Schreiber}, {Schuette}, {Schutz}, {Scott}, {Scott}, {Sellers}, {Sengupta}, {Sergeev}, {Serna}, {Sevigny}, {Shaddock}, {Shahriar}, {Shaltev}, {Shao}, {Shapiro}, {Shawhan}, {Shoemaker}, {Sidery}, {Siemens}, {Sigg}, {Silva}, {Simakov}, {Singer},
  {Singer}, {Singh}, {Sintes}, {Slagmolen}, {Smith}, {Smith}, {Smith}, {Smith-Lefebvre}, {Son}, {Sorazu}, {Souradeep}, {Staley}, {Stebbins}, {Steinke}, {Steinlechner}, {Steinlechner}, {Steinmeyer}, {Stephens}, {Steplewski}, {Stevenson}, {Stone}, {Strain}, {Strigin}, {Sturani}, {Stuver}, {Summerscales}, {Sutton}, {Szczepanczyk}, {Szeifert}, {Talukder}, {Tanner}, {T{\'a}pai}, {Tarabrin}, {Taracchini}, {Taylor}, {Tellez}, {Theeg}, {Thirugnanasambandam}, {Thomas}, {Thomas}, {Thorne}, {Thorne}, {Thrane}, {Tiwari}, {Tomlinson}, {Torres}, {Torrie}, {Traylor}, {Tse}, {Tshilumba}, {Ugolini}, {Unnikrishnan}, {Urban}, {Usman}, {Vahlbruch}, {Vajente}, {Valdes}, {Vallisneri}, {van Veggel}, {Vass}, {Vaulin}, {Vecchio}, {Veitch}, {Veitch}, {Venkateswara}, {Vincent-Finley}, {Vitale}, {Vo}, {Vorvick}, {Vousden}, {Vyatchanin}, {Wade}, {Wade}, {Wade}, {Walker}, {Wallace}, {Walsh}, {Wang}, {Wang}, {Wang}, {Ward}, {Warner}, {Was}, {Weaver}, {Weinert}, {Weinstein}, {Weiss}, {Welborn}, {Wen}, {Wessels}, {Westphal}, {Wette},
  {Whelan}, {Whitcomb}, {White}, {Whiting}, {Wilkinson}, {Williams}, {Williams}, {Williamson}, {Willis}, {Willke}, {Wimmer}, {Winkler}, {Wipf}, {Wittel}, {Woan}, {Worden}, {Xie}, {Yablon}, {Yakushin}, {Yam}, {Yamamoto}, {Yancey}, {Yang}, {Zanolin}, {Zhang}, {Zhang}, {Zhang}, {Zhang}, {Zhao}, {Zhou}, {Zhu}, {Zucker}, {Zuraw}, \& {Zweizig}}]{aasi2015}
{Aasi}, J., {Abbott}, B.~P., {Abbott}, R., {et~al.} 2015, Classical and Quantum Gravity, 32, 074001

\bibitem[{{Abac} {et~al.}(2024){Abac}, {Abbott}, {Abouelfettouh}, {Acernese}, {Ackley}, {Adhicary}, {Adhikari}, {Adhikari}, {Adkins}, {Agarwal}, {Agathos}, {Abchouyeh}, {Aguiar}, {Aguilar}, {Aiello}, {Ain}, {Ajith}, {Ak{\c{c}}ay}, {Akutsu}, {Albanesi}, {Alfaidi}, {Al-Jodah}, {All{\'e}n{\'e}}, {Allocca}, {Al-Shammari}, {Altin}, {Alvarez-Lopez}, {Amato}, {Amez-Droz}, {Amorosi}, {Amra}, {Ananyeva}, {Anderson}, {Anderson}, {Andia}, {Ando}, {Andrade}, {Andres}, {Andr{\'e}s-Carcasona}, {Andri{\'c}}, {Anglin}, {Ansoldi}, {Antelis}, {Antier}, {Aoumi}, {Appavuravther}, {Appert}, {Apple}, {Arai}, {Araya}, {Araya}, {Areeda}, {Argianas}, {Aritomi}, {Armato}, {Arnaud}, {Arogeti}, {Aronson}, {Arun}, {Ashton}, {Aso}, {Assiduo}, {de Souza Melo}, {Aston}, {Astone}, {Attadio}, {Aubin}, {Aultoneal}, {Avallone}, {Azrad}, {Babak}, {Badaracco}, {Badger}, {Bae}, {Bagnasco}, {Bagui}, {Baier}, {Baiotti}, {Bajpai}, {Baka}, {Ball}, {Ballardin}, {Ballmer}, {Banagiri}, {Banerjee}, {Bankar}, {Baral}, {Barayoga}, {Barish}, {Barker},
  {Barneo}, {Barone}, {Barr}, {Barsotti}, {Barsuglia}, {Barta}, {Bartoletti}, {Barton}, {Bartos}, {Basak}, {Basalaev}, {Bassiri}, {Basti}, {Bates}, {Bawaj}, {Baxi}, {Bayley}, {Baylor}, {Baynard}, {Bazzan}, {Bedakihale}, {Beirnaert}, {Bejger}, {Belardinelli}, {Bell}, {Benedetto}, {Benoit}, {Bentara}, {Bentley}, {Ben Yaala}, {Bera}, {Berbel}, {Bergamin}, {Berger}, {Bernuzzi}, {Beroiz}, {Berry}, {Bersanetti}, {Bertolini}, {Betzwieser}, {Beveridge}, {Bevins}, {Bhandare}, {Bhardwaj}, {Bhatt}, {Bhattacharjee}, {Bhaumik}, {Bhowmick}, {Bianchi}, {Bilenko}, {Billingsley}, {Binetti}, {Bini}, {Birnholtz}, {Biscoveanu}, {Bisht}, {Bitossi}, {Bizouard}, {Blackburn}, {Blagg}, {Blair}, {Blair}, {Bobba}, {Bode}, {Boileau}, {Boldrini}, {Bolingbroke}, {Bolliand}, {Bonavena}, {Bondarescu}, {Bondu}, {Bonilla}, {Bonilla}, {Bonino}, {Bonnand}, {Booker}, {Borchers}, {Boschi}, {Bose}, {Bossilkov}, {Boudart}, {Boudon}, {Bozzi}, {Bradaschia}, {Brady}, {Braglia}, {Branch}, {Branchesi}, {Brandt}, {Braun}, {Breschi}, {Briant}, {Brillet},
  {Brinkmann}, {Brockill}, {Brockmueller}, {Brooks}, {Brown}, {Brown}, {Brozzetti}, {Brunett}, {Bruno}, {Bruntz}, {Bryant}, {Bucci}, {Buchanan}, {Bulashenko}, {Bulik}, {Bulten}, {Buonanno}, {Burtnyk}, {Buscicchio}, {Buskulic}, {Buy}, {Byer}, {Cabourn Davies}, {Cabras}, {Cabrita}, {C{\'a}ceres-Barbosa}, {Cadonati}, {Cagnoli}, {Cahillane}, {Bustillo}, {Callister}, {Calloni}, {Camp}, {Canepa}, {Caneva Santoro}, {Cannon}, {Cao}, {Capistran}, {Capocasa}, {Capote}, {Carapella}, {Carbognani}, {Carlassara}, {Carlin}, {Carpinelli}, {Carrillo}, {Carter}, {Carullo}, {Casanueva Diaz}, {Casentini}, {Castro-Lucas}, {Caudill}, {Cavagli{\`a}}, {Cavalieri}, {Cella}, {Cerd{\'a}-Dur{\'a}n}, {Cesarini}, {Chaibi}, {Chakraborty}, {Subrahmanya}, {Chan}, {Chan}, {Chandra}, {Chang}, {Chao}, {Char}, {Charlton}, {Charlton}, {Chassande-Mottin}, {Chatterjee}, {Chatterjee}, {Chatterjee}, {Chattopadhyay}, {Chaturvedi}, {Chaty}, {Chatziioannou}, {Chen}, {Chen}, {Chen}, {Chen}, {Chen}, {Chen}, {Chen}, {Chen}, {Chen}, {Chen}, {Cheng},
  {Chessa}, {Cheung}, {Cheung}, {Chiadini}, {Chiarini}, {Chierici}, {Chincarini}, {Chiofalo}, {Chiummo}, {Chou}, {Choudhary}, {Christensen}, {Chua}, {Chugh}, {Ciani}, {Ciecielag}, {Cie{\'s}lar}, {Cifaldi}, {Ciolfi}, {Clara}, {Clark}, {Clarke}, {Clarke}, {Clearwater}, {Clesse}, {Coccia}, {Codazzo}, {Cohadon}, {Colace}, {Colleoni}, {Collette}, {Collins}, {Colloms}, {Colombo}, {Colpi}, {Compton}, {Connolly}, {Conti}, {Corbitt}, {Cordero-Carri{\'o}n}, {Corezzi}, {Cornish}, {Corsi}, {Cortese}, {Costa}, {Cottingham}, {Coughlin}, {Couineaux}, {Coulon}, {Countryman}, {Coupechoux}, {Couvares}, {Coward}, {Cowart}, {Coyne}, {Craig}, {Creed}, {Creighton}, {Creighton}, {Cremonese}, {Criswell}, {Crockett-Gray}, {Crook}, {Crouch}, {Csizmazia}, {Cudell}, {Cullen}, {Cumming}, {Cuoco}, {Cusinato}, {Dabadie}, {Dal Canton}, {Dall'Osso}, {Dal Pra}, {D{\'a}lya}, {D'Angelo}, {Danilishin}, {D'Antonio}, {Danzmann}, {Darroch}, {Dartez}, {Dasgupta}, {Datta}, {Dattilo}, {Daumas}, {Davari}, {Dave}, {Davenport}, {Davier}, {Davies},
  {Davis}, {Davis}, {Davis}, {Davis}, {Dax}, {de Bolle}, {Deenadayalan}, {Degallaix}, {de Laurentis}, {Del{\'e}glise}, {de Lillo}, {Dell'Aquila}, {Del Pozzo}, {De Marco}, {de Matteis}, {D'Emilio}, {Demos}, {Dent}, {Depasse}, {Depergola}, {de Pietri}, {De Rosa}, {de Rossi}, {Desalvo}, {de Simone}, {Dhani}, {Diab}, {D{\'\i}az}, {di Cesare}, {Dideron}, {Didio}, {Dietrich}, {di Fiore}, {di Fronzo}, {di Giovanni}, {di Girolamo}, {Diksha}, {di Michele}, {Ding}, {di Pace}, {di Palma}, {di Renzo}, {Divyajyoti}, {Dmitriev}, {Doctor}, {Dohmen}, {Doleva}, {Dominguez}, {D'Onofrio}, {Donovan}, {Dooley}, {Dooney}, {Doravari}, {Dorosh}, {Drago}, {Driggers}, {Ducoin}, {Dunn}, {Dupletsa}, {D'Urso}, {Duval}, {Duverne}, {Dwyer}, {Eassa}, {Ebersold}, {Eckhardt}, {Eddolls}, {Edelman}, {Edo}, {Edy}, {Effler}, {Eichholz}, {Einsle}, {Eisenmann}, {Eisenstein}, {Ejlli}, {Eleveld}, {Emma}, {Endo}, {Engl}, {Enloe}, {Errico}, {Essick}, {Estell{\'e}s}, {Estevez}, {Etzel}, {Evans}, {Evstafyeva}, {Ewing}, {Ezquiaga}, {Fabrizi}, {Faedi},
  {Fafone}, {Fairhurst}, {Farah}, {Farr}, {Farr}, {Favaro}, {Favata}, {Fays}, {Fazio}, {Feicht}, {Fejer}, {Felicetti}, {Fenyvesi}, {Ferguson}, {Ferraiuolo}, {Ferrante}, {Ferreira}, {Fidecaro}, {Figura}, {Fiori}, {Fiori}, {Fishbach}, {Fisher}, {Fittipaldi}, {Fiumara}, {Flaminio}, {Fleischer}, {Fleming}, {Floden}, {Foley}, {Fong}, {Font}, {Fornal}, {Forsyth}, {Franceschetti}, {Franchini}, {Frasca}, {Frasconi}, {Mascioli}, {Frei}, {Freise}, {Freitas}, {Frey}, {Frischhertz}, {Fritschel}, {Frolov}, {Fronz{\'e}}, {Fuentes-Garcia}, {Fujii}, {Fujimori}, {Fulda}, {Fyffe}, {Gadre}, {Gair}, {Galaudage}, {Galdi}, {Gallagher}, {Gallardo}, {Gallego}, {Gamba}, {Gamboa}, {Ganapathy}, {Ganguly}, {Garaventa}, {Garc{\'\i}a-Bellido}, {Garc{\'\i}a N{\'u}{\~n}ez}, {Garc{\'\i}a-Quir{\'o}s}, {Gardner}, {Gardner}, {Gargiulo}, {Garron}, {Garufi}, {Gasbarra}, {Gateley}, {Gayathri}, {Gemme}, {Gennai}, {Gennari}, {George}, {George}, {Gerberding}, {Gergely}, {Ghonge}, {Ghosh}, {Ghosh}, {Ghosh}, {Ghosh}, {Ghosh}, {Ghosh}, {Giacoppo},
  {Giaime}, {Giardina}, {Gibson}, {Gibson}, {Gier}, {Giri}, {Gissi}, {Gkaitatzis}, {Glanzer}, {Glotin}, {Godfrey}, {Godwin}, {Goebbels}, {Goetz}, {Golomb}, {Gomez Lopez}, {Goncharov}, {Gong}, {Gonz{\'a}lez}, {Goodarzi}, {Goode}, {Goodwin-Jones}, {Gosselin}, {G{\"o}ttel}, {Gouaty}, {Gould}, {Govorkova}, {Goyal}, {Grace}, {Grado}, {Graham}, {Granados}, {Granata}, {Granata}, {Gras}, {Grassia}, {Gray}, {Gray}, {Gray}, {Greco}, {Green}, {Green}, {Green}, {Gretarsson}, {Gretarsson}, {Griffith}, {Griffiths}, {Griggs}, {Grignani}, {Grimaldi}, {Grimaud}, {Grote}, {Guerra}, {Guetta}, {Guidi}, {Guimaraes}, {Gulati}, {Gulminelli}, {Gunny}, {Guo}, {Guo}, {Guo}, {Gupta}, {Gupta}, {Gupta}, {Gupta}, {Gupta}, {Gupta}, {Gupta}, {Gupte}, {Gurs}, {Gutierrez}, {Guzman}, {H}, {Haba}, {Haberland}, {Haino}, {Hall}, {Hamilton}, {Hammond}, {Han}, {Haney}, {Hanks}, {Hanna}, {Hannam}, {Hannuksela}, {Hanselman}, {Hansen}, {Hanson}, {Harada}, {Hardison}, {Haris}, {Harmark}, {Harms}, {Harry}, {Harry}, {Hart}, {Haskell}, {Haster},
  {Hathaway}, {Haughian}, {Hayakawa}, {Hayama}, {Hayes}, {Heffernan}, {Heidmann}, {Heintze}, {Heinze}, {Heinzel}, {Heitmann}, {Hellman}, {Hello}, {Helmling-Cornell}, {Hemming}, {Henderson-Sapir}, {Hendry}, {Heng}, {Hennes}, {Henshaw}, {Hertog}, {Heurs}, {Hewitt}, {Heyns}, {Higginbotham}, {Hild}, {Hill}, {Himemoto}, {Hirata}, {Hirose}, {Hoang}, {Hochheim}, {Hofman}, {Holland}, {Holley-Bockelmann}, {Holmes}, {Holz}, {Honet}, {Hong}, {Hornung}, {Hoshino}, {Hough}, {Hourihane}, {Howell}, {Hoy}, {Hrishikesh}, {Hsieh}, {Hsiung}, {Hsu}, {Hsu}, {Hu}, {Hu}, {Huang}, {Huang}, {Huddart}, {Hughey}, {Hui}, {Hui}, {Husa}, {Huxford}, {Huynh-Dinh}, {Iampieri}, {Iandolo}, {Ianni}, {Iess}, {Imafuku}, {Inayoshi}, {Inoue}, {Iorio}, {Iqbal}, {Irwin}, {Ishikawa}, {Isi}, {Ismail}, {Itoh}, {Iwanaga}, {Iwaya}, {Iyer}, {Jaberianhamedan}, {Jacquet}, {Jacquet}, {Jadhav}, {Jadhav}, {Jain}, {James}, {James}, {Jamshidi}, {Janquart}, {Janssens}, {Janthalur}, {Jaraba}, {Jaranowski}, {Jaume}, {Javed}, {Jennings}, {Jia}, {Jiang}, {Kubisz},
  {Johanson}, {Johns}, {Johnson}, {Johnson-McDaniel}, {Johnston}, {Johnston}, {Johny}, {Jones}, {Jones}, {Jones}, {Jose}, {Joshi}, {Ju}, {Jung}, {Junker}, {Juste}, {Kajita}, {Kaku}, {Kalaghatgi}, {Kalogera}, {Kamiizumi}, {Kanda}, {Kandhasamy}, {Kang}, {Kanner}, {Kapadia}, {Kapasi}, {Karat}, {Karathanasis}, {Kashyap}, {Kasprzack}, {Kastaun}, {Kato}, {Katsavounidis}, {Katzman}, {Kaushik}, {Kawabe}, {Kawamoto}, {Kazemi}, {Kedia}, {Keitel}, {Kelley-Derzon}, {Kennington}, {Kesharwani}, {Key}, {Khadela}, {Khadka}, {Khalili}, {Khan}, {Khan}, {Khanam}, {Khursheed}, {Khusid}, {Kiendrebeogo}, {Kijbunchoo}, {Kim}, {Kim}, {Kim}, {Kim}, {Kim}, {Kim}, {Kimball}, {Kinley-Hanlon}, {Kinnear}, {Kissel}, {Klimenko}, {Knee}, {Knust}, {Kobayashi}, {Koch}, {Koehlenbeck}, {Koekoek}, {Kohri}, {Kokeyama}, {Koley}, {Kolitsidou}, {Kolstein}, {Komori}, {Kong}, {Kontos}, {Korobko}, {Kossak}, {Kou}, {Koushik}, {Kouvatsos}, {Kovalam}, {Kozak}, {Kranzhoff}, {Kringel}, {Krishnendu}, {Kr{\'o}lak}, {Kruska}, {Kuehn}, {Kuijer}, {Kulkarni},
  {Ramamohan}, {Kumar}, {Kumar}, {Kumar}, {Kumar}, {Kumar}, {Kume}, {Kuns}, {Kuntimaddi}, {Kuroyanagi}, {Kurth}, {Kuwahara}, {Kwak}, {Kwan}, {Kwok}, {Lacaille}, {Lagabbe}, {Laghi}, {Lai}, {Laity}, {Lakkis}, {Lalande}, {Lalleman}, {Lalremruati}, {Landry}, {Landry}, {Lane}, {Lang}, {Lange}, {Lantz}, {La Rana}, {La Rosa}, {Lartaux-Vollard}, {Lasky}, {Lawrence}, {Lawrence}, {Laxen}, {Lazzarini}, {Lazzaro}, {Leaci}, {Lecoeuche}, {Lee}, {Lee}, {Lee}, {Lee}, {Lee}, {Lee}, {Lee}, {Legred}, {Lehmann}, {Lehner}, {Le Jean}, {Lema{\^\i}tre}, {Lenti}, {Leonardi}, {Lequime}, {Leroy}, {Lesovsky}, {Letendre}, {Lethuillier}, {Levin}, {Levin}, {Leyde}, {Li}, {Li}, {Li}, {Li}, {Li}, {Lihos}, {Lin}, {Lin}, {Lin}, {Lin}, {Lin}, {Lin}, {Lin}, {Linde}, {Linker}, {Littenberg}, {Liu}, {Liu}, {Liu}, {Villarreal}, {Llobera-Querol}, {Lo}, {Locquet}, {London}, {Longo}, {Lopez}, {Lopez Portilla}, {Lorenzini}, {Lorenzo-Medina}, {Loriette}, {Lormand}, {Losurdo}, {}, {Lough}, {Loughlin}, {Lousto}, {Lowry}, {Lu}, {L{\"u}ck}, {Lumaca},
  {Lundgren}, {Lussier}, {Ma}, {Ma}, {Ma'Arif}, {Macas}, {Macedo}, {Macinnis}, {Maciy}, {MacLeod}, {MacMillan}, {Macquet}, {Macri}, {Maeda}, {Maenaut}, {Hernandez}, {Magare}, {Magazz{\`u}}, {Magee}, {Maggio}, {Maggiore}, {Magnozzi}, {Mahesh}, {Mahesh}, {Maini}, {Majhi}, {Majorana}, {Makarem}, {Makelele}, {Malaquias-Reis}, {Mali}, {Maliakal}, {Malik}, {Man}, {Mandic}, {Mangano}, {Mannix}, {Mansell}, {Mansingh}, {Manske}, {Mantovani}, {Mapelli}, {Marchesoni}, {Mar{\'\i}n Pina}, {Marion}, {M{\'a}rka}, {M{\'a}rka}, {Markosyan}, {Markowitz}, {Maros}, {Marsat}, {Martelli}, {Martin}, {Martin}, {Martinez}, {Martinez}, {Martinez}, {Martini}, {Martinovic}, {Martins}, {Martynov}, {Marx}, {Massaro}, {Masserot}, {Masso-Reid}, {Mastrodicasa}, {Mastrogiovanni}, {Matcovich}, {Matiushechkina}, {Matsuyama}, {Mavalvala}, {Maxwell}, {McCarrol}, {McCarthy}, {McClelland}, {McCormick}, {McCuller}, {McEachin}, {McElhenny}, {McGhee}, {McGinn}, {McGowan}, {McIver}, {McLeod}, {McRae}, {Meacher}, {Meijer}, {Melatos}, {Mellaerts},
  {Menendez-Vazquez}, {Menoni}, {Mera}, {Mercer}, {Mereni}, {Merfeld}, {Merilh}, {M{\'e}rou}, {Merritt}, {Merzougui}, {Messenger}, {Messick}, {Meyer-Conde}, {Meylahn}, {Mhaske}, {Miani}, {Miao}, {Michaloliakos}, {Michel}, {Michimura}, {Middleton}, {Miller}, {Miller}, {Millhouse}, {Milotti}, {Milotti}, {Minenkov}, {Mio}, {Mir}, {Mirasola}, {Miravet-Ten{\'e}s}, {Miritescu}, {Mishra}, {Mishra}, {Mishra}, {Mishra}, {Mitchell}, {Mitchell}, {Mitra}, {Mitrofanov}, {Mittleman}, {Miyakawa}, {Miyamoto}, {Miyoki}, {Mo}, {Mobilia}, {Mohapatra}, {Mohite}, {Molina-Ruiz}, {Mondal}, {Mondin}, {Montani}, {Moore}, {Moraru}, {More}, {More}, {Moreno}, {Morgan}, {Morisaki}, {Moriwaki}, {Morras}, {Moscatello}, {Mourier}, {Mours}, {Mow-Lowry}, {Muciaccia}, {Mukherjee}, {Mukherjee}, {Mukherjee}, {Mukherjee}, {Mukherjee}, {Mukherjee}, {Mukund}, {Mullavey}, {Munch}, {Mundi}, {Mungioli}, {Oberg}, {Murakami}, {Murakoshi}, {Murray}, {Muusse}, {Nabari}, {Nadji}, {Nagar}, {Nagarajan}, {Nagler}, {Nakagaki}, {Nakamura}, {Nakano}, {Nakano},
  {Nandi}, {Napolano}, {Narayan}, {Nardecchia}, {Narikawa}, {Narola}, {Naticchioni}, {Nayak}, {Neilson}, {Nelson}, {Nelson}, {Nery}, {Neunzert}, {Ng}, {Quynh}, {Nichols}, {Nielsen}, {Nieradka}, {Niko}, {Nishino}, {Nishizawa}, {Nissanke}, {Nitoglia}, {Niu}, {Nocera}, {Norman}, {North}, {Novak}, {Nu{\~n}o Siles}, {Nuttall}, {Obayashi}, {Oberling}, {O'Dell}, {Oertel}, {Offermans}, {Oganesyan}, {Oh}, {Oh}, {O'Hanlon}, {Ohashi}, {Ohkawa}, {Ohme}, {Oliveira}, {Oliveri}, {O'Neal}, {Oohara}, {O'Reilly}, {Ormsby}, {Orselli}, {O'Shaughnessy}, {O'Shea}, {Oshima}, {Oshino}, {Ossokine}, {Osthelder}, {Ota}, {Ottaway}, {Ouzriat}, {Overmier}, {Owen}, {Pace}, {Pagano}, {Page}, {Pai}, {Pal}, {Pal}, {Palaia}, {P{\'a}lfi}, {Palma}, {Palomba}, {Palud}, {Pan}, {Pan}, {Pan}, {Panai}, {Panda}, {Pandey}, {Panebianco}, {Pang}, {Pannarale}, {Pannone}, {Pant}, {Panther}, {Paoletti}, {Paolone}, {Papalexakis}, {Papalini}, {Papigkiotis}, {Paquis}, {Parisi}, {Park}, {Park}, {Parker}, {Pascale}, {Pascucci}, {Pasqualetti}, {Passaquieti},
  {Passenger}, {Passuello}, {Patane}, {Pathak}, {Pathak}, {Patra}, {Patricelli}, {Patron}, {Paul}, {Paul}, {Payne}, {Pearce}, {Pedraza}, {Pegna}, {Pele}, {Arellano}, {Penn}, {Penuliar}, {Perego}, {Pereira}, {Perez}, {P{\'e}rigois}, {Perna}, {Perreca}, {Perret}, {Perri{\`e}s}, {Perry}, {Pesios}, {Petracca}, {Petrillo}, {Pfeiffer}, {Pham}, {Pham}, {Phukon}, {Phurailatpam}, {Piarulli}, {Piccari}, {Piccinni}, {Pichot}, {Piendibene}, {Piergiovanni}, {Pierini}, {Pierra}, {Pierro}, {Pietrzak}, {Pillas}, {Pilo}, {Pinard}, {Pinto}, {Pinto}, {Piotrzkowski}, {Pirello}, {Pitkin}, {Placidi}, {Placidi}, {Planas}, {Plastino}, {Poggiani}, {Polini}, {Pompili}, {Poon}, {Porcelli}, {Porter}, {Posnansky}, {Poulton}, {Powell}, {Pracchia}, {Pradhan}, {Pradier}, {Prajapati}, {Prasai}, {Prasanna}, {Prasia}, {Pratten}, {Principe}, {Principe}, {Prodi}, {Prokhorov}, {Prosposito}, {Puecher}, {Pullin}, {Punturo}, {Puppo}, {P{\"u}rrer}, {Qi}, {Qin}, {Qu{\'e}m{\'e}ner}, {Quetschke}, {Quigley}, {Quinonez}, {Raab}, {Raabith}, {Raaijmakers},
  {Raja}, {Rajan}, {Rajbhandari}, {Ramirez}, {Vidal}, {Ramos-Buades}, {Rana}, {Ranjan}, {Ransom}, {Rapagnani}, {Ratto}, {Rawat}, {Ray}, {Raymond}, {Razzano}, {Read}, {Payo}, {Regimbau}, {Rei}, {Reid}, {Reitze}, {Relton}, {Renzini}, {Rettegno}, {Revenu}, {Reyes}, {Rezaei}, {Ricci}, {Ricci}, {Ricciardone}, {Richardson}, {Richardson}, {Rijal}, {Riles}, {Riley}, {Rinaldi}, {Rittmeyer}, {Robertson}, {Robinet}, {Robinson}, {Rocchi}, {Rolland}, {Rollins}, {Romano}, {Romano}, {Romero}, {Romero-Shaw}, {Romie}, {Ronchini}, {Roocke}, {Rosa}, {Rosauer}, {Rose}, {Rosi{\'n}ska}, {Ross}, {Rossello}, {Rowan}, {Roy}, {Roy}, {Rozza}, {Ruggi}, {Ruhama}, {Morales}, {Ruiz-Rocha}, {Sachdev}, {Sadecki}, {Sadiq}, {Saffarieh}, {Sah}, {Saha}, {Saha}, {Sainrat}, {Menon}, {Sakai}, {Sakellariadou}, {Sakon}, {Salafia}, {Salces-Carcoba}, {Salconi}, {Saleem}, {Salemi}, {Sall{\'e}}, {Salvador}, {Sanchez}, {Sanchez}, {Sanchez}, {Sanchez}, {Sanchis-Gual}, {Sanders}, {S{\"a}nger}, {Santoliquido}, {Saravanan}, {Sarin}, {Sasaoka}, {Sasli},
  {Sassi}, {Sassolas}, {Satari}, {Sathyaprakash}, {Sato}, {Sato}, {Sauter}, {Savage}, {Sawada}, {Sawant}, {Sayah}, {Scacco}, {Schaetzl}, {Scheel}, {Schiebelbein}, {Schiworski}, {Schmidt}, {Schmidt}, {Schnabel}, {Schneewind}, {Schofield}, {Schouteden}, {Schulte}, {Schutz}, {Schwartz}, {Scialpi}, {Scott}, {Scott}, {Seetharamu}, {Seglar-Arroyo}, {Sekiguchi}, {Sellers}, {Sengupta}, {Sentenac}, {Seo}, {Seo}, {Sequino}, {Serra}, {Servignat}, {Sevrin}, {Shaffer}, {Shah}, {Shaikh}, {Shao}, {Sharma}, {Sharma}, {Sharma-Chaudhary}, {Shaw}, {Shawhan}, {Shcheblanov}, {Sheridan}, {Shikano}, {Shikauchi}, {Shimode}, {Shinkai}, {Shiota}, {Shoemaker}, {Shoemaker}, {Short}, {Shyamsundar}, {Sider}, {Siegel}, {Sieniawska}, {Sigg}, {Silenzi}, {Simmonds}, {Singer}, {Singh}, {Singh}, {Singh}, {Singh}, {Singha}, {Sintes}, {Sipala}, {Skliris}, {Slagmolen}, {Slaven-Blair}, {Smetana}, {Smith}, {Smith}, {Smith}, {Smith}, {Soldateschi}, {Somiya}, {Song}, {Soni}, {Soni}, {Sordini}, {Sorrentino}, {Sorrentino}, {Sotani}, {Soulard},
  {Southgate}, {Spagnuolo}, {Spencer}, {Spera}, {Spinicelli}, {Spoon}, {Sprague}, {Srivastava}, {Stachurski}, {Steer}, {Steinlechner}, {Steinlechner}, {Stergioulas}, {Stevens}, {Stevenson}, {Stpierre}, {Stratta}, {Strong}, {Strunk}, {Sturani}, {Stuver}, {Suchenek}, {Sudhagar}, {Sueltmann}, {Suleiman}, {Sullivan}, {Sun}, {Sunil}, {Suresh}, {Sutton}, {Suzuki}, {Suzuki}, {Swinkels}, {Syx}, {Szczepa{\'n}czyk}, {Szewczyk}, {Tacca}, {Tagoshi}, {Tait}, {Takahashi}, {Takahashi}, {Takamori}, {Takase}, {Takatani}, {Takeda}, {Takeshita}, {Talbot}, {Tamaki}, {Tamanini}, {Tanabe}, {Tanaka}, {Tanaka}, {Tanaka}, {Tang}, {Tanioka}, {Tanner}, {Tao}, {Tapia}, {San Mart{\'\i}n}, {Tarafder}, {Taranto}, {Taruya}, {Tasson}, {Teloi}, {Tenorio}, {Themann}, {Theodoropoulos}, {Thirugnanasambandam}, {Thomas}, {Thomas}, {Thomas}, {Thompson}, {Thondapu}, {Thorne}, {Thrane}, {Tissino}, {Tiwari}, {Tiwari}, {Tiwari}, {Tiwari}, {Todd}, {Toivonen}, {Toland}, {Tolley}, {Tomaru}, {Tomita}, {Tomura}, {Tong}, {Tong-Yu}, {Toriyama}, {Toropov},
  {Torres-Forn{\'e}}, {Torrie}, {Toscani}, {E Melo}, {Tournefier}, {Trapananti}, {Travasso}, {Traylor}, {Trevor}, {Tringali}, {Tripathee}, {Troian}, {Troiano}, {Trovato}, {Trozzo}, {Trudeau}, {Tsang}, {Tso}, {Tsuchida}, {Tsukada}, {Tsutsui}, {Turbang}, {Turconi}, {Turski}, {Ubach}, {Uchikata}, {Uchiyama}, {Udall}, {Uehara}, {Uematsu}, {Ueno}, {Ueno}, {Undheim}, {Ushiba}, {Vacatello}, {Vahlbruch}, {Vaidya}, {Vajente}, {Vajpeyi}, {Valdes}, {Valencia}, {Valentini}, {Vallejo-Pe{\~n}a}, {Vallero}, {Valsan}, {van Bakel}, {van Beuzekom}, {van Dael}, {van den Brand}, {Broeck}, {Vander-Hyde}, {van der Sluys}, {van de Walle}, {van Dongen}, {Vandra}, {van Haevermaet}, {van Heijningen}, {van Hove}, {Vankeuren}, {Vanosky}, {van Putten}, {van Ranst}, {van Remortel}, {Vardaro}, {Vargas}, {Varghese}, {Varma}, {Vas{\'u}th}, {Vecchio}, {Vedovato}, {Veitch}, {Veitch}, {Venikoudis}, {Venneberg}, {Verdier}, {Verkindt}, {Verma}, {Verma}, {Verma}, {Vermeulen}, {Vetrano}, {Veutro}, {Vibhute}, {Vicer{\'e}}, {Vidyant}, {Viets},
  {Vijaykumar}, {Vilkha}, {Villa-Ortega}, {Vincent}, {Vinet}, {Viret}, {Virtuoso}, {Vitale}, {Vives}, {Vocca}, {Voigt}, {von Reis}, {von Wrangel}, {Vyatchanin}, {Wade}, {Wade}, {Wagner}, {Wajid}, {Walker}, {Wallace}, {Wallace}, {Wang}, {Wang}, {Wang}, {Wang}, {Waratkar}, {Warner}, {Was}, {Washimi}, {Washington}, {Watarai}, {Wayt}, {Weaver}, {Weaver}, {Weaving}, {Webster}, {Weinert}, {Weinstein}, {Weiss}, {Wellmann}, {Wen}, {We{\ss}els}, {Wette}, {Whelan}, {Whiting}, {Whittle}, {Wildberger}, {Wilk}, {Wilken}, {Wilkin}, {Willadsen}, {Willetts}, {Williams}, {Williams}, {Williams}, {Willis}, {Willke}, {Wils}, {Winterflood}, {Wipf}, {Woan}, {Woehler}, {Wofford}, {Wolfe}, {Wong}, {Wong}, {Wong}, {Wright}, {Wright}, {Wu}, {Wu}, {Wu}, {Wuchner}, {Wysocki}, {Xu}, {Xu}, {Yadav}, {Yamamoto}, {Yamamoto}, {Yamamoto}, {Yamamoto}, {Yamamura}, {Yamazaki}, {Yan}, {Yan}, {Yang}, {Yang}, {Yang}, {Yang}, {Yarbrough}, {Yasui}, {Yeh}, {Yelikar}, {Yin}, {Yokoyama}, {Yokozawa}, {Yoo}, {Yu}, {Yuan}, {Yuzurihara}, {Zadro{\.z}ny},
  {Zanolin}, {Zeeshan}, {Zelenova}, {Zendri}, {Zeoli}, {Zerrad}, {Zevin}, {Zhang}, {Zhang}, {Zhang}, {Zhang}, {Zhang}, {Zhao}, {Zhao}, {Zhao}, {Zheng}, {Zhong}, {Zhou}, {Zhu}, {Zhu}, {Zimmerman}, {Zucker}, {Zweizig}, {Ligo Scientific Collaboration}, {VIRGO Collaboration}, \& {Kagra Collaboration}}]{GW230529}
{Abac}, A.~G., {Abbott}, R., {Abouelfettouh}, I., {et~al.} 2024, \apjl, 970, L34

\bibitem[{{Abbott} {et~al.}(2021){Abbott}, {Abbott}, {Abraham}, {Acernese}, {Ackley}, {Adams}, {Adams}, {Adhikari}, {Adya}, {Affeldt}, {Agarwal}, {Agathos}, {Agatsuma}, {Aggarwal}, {Aguiar}, {Aiello}, {Ain}, {Ajith}, {Akutsu}, {Aleman}, {Allen}, {Allocca}, {Altin}, {Amato}, {Anand}, {Ananyeva}, {Anderson}, {Anderson}, {Ando}, {Angelova}, {Ansoldi}, {Antelis}, {Antier}, {Appert}, {Arai}, {Arai}, {Arai}, {Araki}, {Araya}, {Araya}, {Areeda}, {Ar{\`e}ne}, {Aritomi}, {Arnaud}, {Aronson}, {Arun}, {Asada}, {Asali}, {Ashton}, {Aso}, {Aston}, {Astone}, {Aubin}, {Aufmuth}, {Aultoneal}, {Austin}, {Babak}, {Badaracco}, {Bader}, {Bae}, {Bae}, {Baer}, {Bagnasco}, {Bai}, {Baiotti}, {Baird}, {Bajpai}, {Ball}, {Ballardin}, {Ballmer}, {Bals}, {Balsamo}, {Baltus}, {Banagiri}, {Bankar}, {Bankar}, {Barayoga}, {Barbieri}, {Barish}, {Barker}, {Barneo}, {Barone}, {Barr}, {Barsotti}, {Barsuglia}, {Barta}, {Bartlett}, {Barton}, {Bartos}, {Bassiri}, {Basti}, {Bawaj}, {Bayley}, {Baylor}, {Bazzan}, {B{\'e}csy}, {Bedakihale}, {Bejger},
  {Belahcene}, {Benedetto}, {Beniwal}, {Benjamin}, {Benkel}, {Bennett}, {Bentley}, {Benyaala}, {Bergamin}, {Berger}, {Bernuzzi}, {Berry}, {Bersanetti}, {Bertolini}, {Betzwieser}, {Bhandare}, {Bhandari}, {Bhattacharjee}, {Bhaumik}, {Bidler}, {Bilenko}, {Billingsley}, {Birney}, {Birnholtz}, {Biscans}, {Bischi}, {Biscoveanu}, {Bisht}, {Biswas}, {Bitossi}, {Bizouard}, {Blackburn}, {Blackman}, {Blair}, {Blair}, {Blair}, {Bobba}, {Bode}, {Boer}, {Bogaert}, {Boldrini}, {Bondu}, {Bonilla}, {Bonnand}, {Booker}, {Boom}, {Bork}, {Boschi}, {Bose}, {Bose}, {Bossilkov}, {Boudart}, {Bouffanais}, {Bozzi}, {Bradaschia}, {Brady}, {Bramley}, {Branch}, {Branchesi}, {Brau}, {Breschi}, {Briant}, {Briggs}, {Brillet}, {Brinkmann}, {Brockill}, {Brooks}, {Brooks}, {Brown}, {Brunett}, {Bruno}, {Bruntz}, {Bryant}, {Buikema}, {Bulik}, {Bulten}, {Buonanno}, {Buscicchio}, {Buskulic}, {Byer}, {Cadonati}, {Caesar}, {Cagnoli}, {Cahillane}, {Cain}, {Calder{\'o}n Bustillo}, {Callaghan}, {Callister}, {Calloni}, {Camp}, {Canepa},
  {Cannavacciuolo}, {Cannon}, {Cao}, {Cao}, {Cao}, {Capocasa}, {Capote}, {Carapella}, {Carbognani}, {Carlin}, {Carney}, {Carpinelli}, {Carullo}, {Carver}, {Casanueva Diaz}, {Casentini}, {Castaldi}, {Caudill}, {Cavagli{\`a}}, {Cavalier}, {Cavalieri}, {Cella}, {Cerd{\'a}-Dur{\'a}n}, {Cesarini}, {Chaibi}, {Chakravarti}, {Champion}, {Chan}, {Chan}, {Chan}, {Chan}, {Chandra}, {Chanial}, {Chao}, {Charlton}, {Chase}, {Chassande-Mottin}, {Chatterjee}, {Chaturvedi}, {Chatziioannou}, {Chen}, {Chen}, {Chen}, {Chen}, {Chen}, {Chen}, {Chen}, {Chen}, {Chen}, {Cheng}, {Cheong}, {Cheung}, {Chia}, {Chiadini}, {Chiang}, {Chierici}, {Chincarini}, {Chiofalo}, {Chiummo}, {Cho}, {Cho}, {Choate}, {Choudhary}, {Choudhary}, {Christensen}, {Chu}, {Chu}, {Chu}, {Chua}, {Chung}, {Ciani}, {Ciecielag}, {Cie{\'s}lar}, {Cifaldi}, {Ciobanu}, {Ciolfi}, {Cipriano}, {Cirone}, {Clara}, {Clark}, {Clark}, {Clarke}, {Clearwater}, {Clesse}, {Cleva}, {Coccia}, {Cohadon}, {Cohen}, {Cohen}, {Colleoni}, {Collette}, {Colpi}, {Compton}, {Constancio},
  {Conti}, {Cooper}, {Corban}, {Corbitt}, {Cordero-Carri{\'o}n}, {Corezzi}, {Corley}, {Cornish}, {Corre}, {Corsi}, {Cortese}, {Costa}, {Cotesta}, {Coughlin}, {Coughlin}, {Coulon}, {Countryman}, {Cousins}, {Couvares}, {Covas}, {Coward}, {Cowart}, {Coyne}, {Coyne}, {Creighton}, {Creighton}, {Criswell}, {Croquette}, {Crowder}, {Cudell}, {Cullen}, {Cumming}, {Cummings}, {Cuoco}, {Cury{\l}o}, {Dal Canton}, {D{\'a}lya}, {Dana}, {Daneshgaranbajastani}, {D'Angelo}, {Danilishin}, {D'Antonio}, {Danzmann}, {Darsow-Fromm}, {Dasgupta}, {Datrier}, {Dattilo}, {Dave}, {Davier}, {Davies}, {Davis}, {Daw}, {Dean}, {Debra}, {Deenadayalan}, {Degallaix}, {de Laurentis}, {Del{\'e}glise}, {Del Favero}, {de Lillo}, {de Lillo}, {Del Pozzo}, {Demarchi}, {de Matteis}, {D'Emilio}, {Demos}, {Dent}, {Depasse}, {de Pietri}, {De Rosa}, {de Rossi}, {Desalvo}, {de Simone}, {Dhurandhar}, {D{\'\i}az}, {Diaz-Ortiz}, {Didio}, {Dietrich}, {di Fiore}, {di Fronzo}, {di Giorgio}, {di Giovanni}, {di Girolamo}, {di Lieto}, {Ding}, {di Pace}, {di Palma},
  {di Renzo}, {Divakarla}, {Dmitriev}, {Doctor}, {D'Onofrio}, {Donovan}, {Dooley}, {Doravari}, {Dorrington}, {Drago}, {Driggers}, {Drori}, {Du}, {Ducoin}, {Dupej}, {Durante}, {D'Urso}, {Duverne}, {Dwyer}, {Easter}, {Ebersold}, {Eddolls}, {Edelman}, {Edo}, {Edy}, {Effler}, {Eguchi}, {Eichholz}, {Eikenberry}, {Eisenmann}, {Eisenstein}, {Ejlli}, {Enomoto}, {Errico}, {Essick}, {Estell{\'e}s}, {Estevez}, {Etienne}, {Etzel}, {Evans}, {Evans}, {Ewing}, {Fafone}, {Fair}, {Fairhurst}, {Fan}, {Farah}, {Farinon}, {Farr}, {Farr}, {Farrow}, {Fauchon-Jones}, {Favata}, {Fays}, {Fazio}, {Feicht}, {Fejer}, {Feng}, {Fenyvesi}, {Ferguson}, {Fernandez-Galiana}, {Ferrante}, {Ferreira}, {Fidecaro}, {Figura}, {Fiori}, {Fishbach}, {Fisher}, {Fittipaldi}, {Fiumara}, {Flaminio}, {Floden}, {Flynn}, {Fong}, {Font}, {Fornal}, {Forsyth}, {Franke}, {Frasca}, {Frasconi}, {Frederick}, {Frei}, {Freise}, {Frey}, {Fritschel}, {Frolov}, {Fronz{\'e}}, {Fujii}, {Fujikawa}, {Fukunaga}, {Fukushima}, {Fulda}, {Fyffe}, {Gabbard}, {Gadre}, {Gaebel},
  {Gair}, {Gais}, {Galaudage}, {Gamba}, {Ganapathy}, {Ganguly}, {Gao}, {Gaonkar}, {Garaventa}, {Garc{\'\i}a-N{\'u}{\~n}ez}, {Garc{\'\i}a-Quir{\'o}s}, {Garufi}, {Gateley}, {Gaudio}, {Gayathri}, {Ge}, {Gemme}, {Gennai}, {George}, {Gergely}, {Gewecke}, {Ghonge}, {Ghosh}, {Ghosh}, {Ghosh}, {Ghosh}, {Ghosh}, {Giacomazzo}, {Giacoppo}, {Giaime}, {Giardina}, {Gibson}, {Gier}, {Giesler}, {Giri}, {Gissi}, {Glanzer}, {Gleckl}, {Godwin}, {Goetz}, {Goetz}, {Gohlke}, {Goncharov}, {Gonz{\'a}lez}, {Gopakumar}, {Gosselin}, {Gouaty}, {Grace}, {Grado}, {Granata}, {Granata}, {Grant}, {Gras}, {Grassia}, {Gray}, {Gray}, {Greco}, {Green}, {Green}, {Gretarsson}, {Gretarsson}, {Griffith}, {Griffiths}, {Griggs}, {Grignani}, {Grimaldi}, {Grimes}, {Grimm}, {Grote}, {Grunewald}, {Gruning}, {Guerrero}, {Guidi}, {Guimaraes}, {Guix{\'e}}, {Gulati}, {Guo}, {Guo}, {Gupta}, {Gupta}, {Gupta}, {Gustafson}, {Gustafson}, {Guzman}, {Ha}, {Haegel}, {Hagiwara}, {Haino}, {Halim}, {Hall}, {Hamilton}, {Hammond}, {Han}, {Haney}, {Hanks}, {Hanna},
  {Hannam}, {Hannuksela}, {Hansen}, {Hansen}, {Hanson}, {Harder}, {Hardwick}, {Haris}, {Harms}, {Harry}, {Harry}, {Hartwig}, {Hasegawa}, {Haskell}, {Hasskew}, {Haster}, {Hattori}, {Haughian}, {Hayakawa}, {Hayama}, {Hayes}, {Healy}, {Heidmann}, {Heintze}, {Heinze}, {Heinzel}, {Heitmann}, {Hellman}, {Hello}, {Helmling-Cornell}, {Hemming}, {Hendry}, {Heng}, {Hennes}, {Hennig}, {Hennig}, {Hernandez Vivanco}, {Heurs}, {Hild}, {Hill}, {Himemoto}, {Hinderer}, {Hines}, {Hiranuma}, {Hirata}, {Hirose}, {Ho}, {Hochheim}, {Hofman}, {Hohmann}, {Holgado}, {Holland}, {Hollows}, {Holmes}, {Holt}, {Holz}, {Hong}, {Hopkins}, {Hough}, {Howell}, {Hoy}, {Hoyland}, {Hreibi}, {Hsieh}, {Hsu}, {Huang}, {Huang}, {Huang}, {Huang}, {Huang}, {Huang}, {H{\"u}bner}, {Huddart}, {Huerta}, {Hughey}, {Hui}, {Hui}, {Husa}, {Huttner}, {Huxford}, {Huynh-Dinh}, {Ide}, {Idzkowski}, {Iess}, {Ikenoue}, {Imam}, {Inayoshi}, {Inchauspe}, {Ingram}, {Inoue}, {Intini}, {Ioka}, {Isi}, {Isleif}, {Ito}, {Itoh}, {Iyer}, {Izumi}, {Jaberianhamedan}, {Jacqmin},
  {Jadhav}, {Jadhav}, {James}, {Jan}, {Jani}, {Janssens}, {Janthalur}, {Jaranowski}, {Jariwala}, {Jaume}, {Jenkins}, {Jeon}, {Jeunon}, {Jia}, {Jiang}, {Jin}, {Johns}, {Jones}, {Jones}, {Jones}, {Jones}, {Jones}, {Jonker}, {Ju}, {Jung}, {Jung}, {Junker}, {Kaihotsu}, {Kajita}, {Kakizaki}, {Kalaghatgi}, {Kalogera}, {Kamai}, {Kamiizumi}, {Kanda}, {Kandhasamy}, {Kang}, {Kanner}, {Kao}, {Kapadia}, {Kapasi}, {Karat}, {Karathanasis}, {Karki}, {Kashyap}, {Kasprzack}, {Kastaun}, {Katsanevas}, {Katsavounidis}, {Katzman}, {Kaur}, {Kawabe}, {Kawaguchi}, {Kawai}, {Kawasaki}, {K{\'e}f{\'e}lian}, {Keitel}, {Key}, {Khadka}, {Khalili}, {Khan}, {Khan}, {Khazanov}, {Khetan}, {Khursheed}, {Kijbunchoo}, {Kim}, {Kim}, {Kim}, {Kim}, {Kim}, {Kim}, {Kimball}, {Kimura}, {King}, {Kinley-Hanlon}, {Kirchhoff}, {Kissel}, {Kita}, {Kitazawa}, {Kleybolte}, {Klimenko}, {Knee}, {Knowles}, {Knyazev}, {Koch}, {Koekoek}, {Kojima}, {Kokeyama}, {Koley}, {Kolitsidou}, {Kolstein}, {Komori}, {Kondrashov}, {Kong}, {Kontos}, {Koper}, {Korobko}, {Kotake},
  {Kovalam}, {Kozak}, {Kozakai}, {Kozu}, {Kringel}, {Krishnendu}, {Kr{\'o}lak}, {Kuehn}, {Kuei}, {Kumar}, {Kumar}, {Kumar}, {Kumar}, {Kume}, {Kuns}, {Kuo}, {Kuo}, {Kuromiya}, {Kuroyanagi}, {Kusayanagi}, {Kwak}, {Kwang}, {Laghi}, {Lalande}, {Lam}, {Lamberts}, {Landry}, {Landry}, {Lane}, {Lang}, {Lange}, {Lantz}, {La Rosa}, {Lartaux-Vollard}, {Lasky}, {Laxen}, {Lazzarini}, {Lazzaro}, {Leaci}, {Leavey}, {Lecoeuche}, {Lee}, {Lee}, {Lee}, {Lee}, {Lee}, {Lee}, {Lehmann}, {Lema{\^\i}tre}, {Leon}, {Leonardi}, {Leroy}, {Letendre}, {Levin}, {Leviton}, {Li}, {Li}, {Li}, {Li}, {Li}, {Li}, {Lin}, {Lin}, {Lin}, {Lin}, {Lin}, {Linde}, {Linker}, {Linley}, {Littenberg}, {Liu}, {Liu}, {Liu}, {Liu}, {Llorens-Monteagudo}, {Lo}, {Lockwood}, {Lollie}, {London}, {Longo}, {Lopez}, {Lorenzini}, {Loriette}, {Lormand}, {Losurdo}, {Lough}, {Lousto}, {Lovelace}, {L{\"u}ck}, {Lumaca}, {Lundgren}, {Luo}, {Macas}, {Macinnis}, {MacLeod}, {MacMillan}, {Macquet}, {Maga{\~n}a Hernandez}, {Maga{\~n}a-Sandoval}, {Magazz{\`u}}, {Magee},
  {Maggiore}, {Majorana}, {Makarem}, {Maksimovic}, {Maliakal}, {Malik}, {Man}, {Mandic}, {Mangano}, {Mango}, {Mansell}, {Manske}, {Mantovani}, {Mapelli}, {Marchesoni}, {Marchio}, {Marion}, {Mark}, {M{\'a}rka}, {M{\'a}rka}, {Markakis}, {Markosyan}, {Markowitz}, {Maros}, {Marquina}, {Marsat}, {Martelli}, {Martin}, {Martin}, {Martinez}, {Martinez}, {Martinovic}, {Martynov}, {Marx}, {Masalehdan}, {Mason}, {Massera}, {Masserot}, {Massinger}, {Masso-Reid}, {Mastrogiovanni}, {Matas}, {Mateu-Lucena}, {Matichard}, {Matiushechkina}, {Mavalvala}, {McCann}, {McCarthy}, {McClelland}, {McClincy}, {McCormick}, {McCuller}, {McGhee}, {McGuire}, {McIsaac}, {McIver}, {McManus}, {McRae}, {McWilliams}, {Meacher}, {Mehmet}, {Mehta}, {Melatos}, {Melchor}, {Mendell}, {Menendez-Vazquez}, {Menoni}, {Mercer}, {Mereni}, {Merfeld}, {Merilh}, {Merritt}, {Merzougui}, {Meshkov}, {Messenger}, {Messick}, {Meyers}, {Meylahn}, {Mhaske}, {Miani}, {Miao}, {Michaloliakos}, {Michel}, {Michimura}, {Middleton}, {Milano}, {Miller}, {Millhouse},
  {Mills}, {Milotti}, {Milovich-Goff}, {Minazzoli}, {Minenkov}, {Mio}, {Mir}, {Mishkin}, {Mishra}, {Mishra}, {Mistry}, {Mitra}, {Mitrofanov}, {Mitselmakher}, {Mittleman}, {Miyakawa}, {Miyamoto}, {Miyazaki}, {Miyo}, {Miyoki}, {Mo}, {Mogushi}, {Mohapatra}, {Mohite}, {Molina}, {Molina-Ruiz}, {Mondin}, {Montani}, {Moore}, {Moraru}, {Morawski}, {More}, {Moreno}, {Moreno}, {Mori}, {Morisaki}, {Moriwaki}, {Mours}, {Mow-Lowry}, {Mozzon}, {Muciaccia}, {Mukherjee}, {Mukherjee}, {Mukherjee}, {Mukherjee}, {Mukund}, {Mullavey}, {Munch}, {Mu{\~n}iz}, {Murray}, {Musenich}, {Nadji}, {Nagano}, {Nagano}, {Nagar}, {Nakamura}, {Nakano}, {Nakano}, {Nakashima}, {Nakayama}, {Nardecchia}, {Narikawa}, {Naticchioni}, {Nayak}, {Nayak}, {Negishi}, {Neil}, {Neilson}, {Nelemans}, {Nelson}, {Nery}, {Neunzert}, {Ng}, {Ng}, {Nguyen}, {Nguyen}, {Nguyen}, {Nguyen Quynh}, {Ni}, {Nichols}, {Nishizawa}, {Nissanke}, {Nocera}, {Noh}, {Norman}, {North}, {Nozaki}, {Nuttall}, {Oberling}, {O'Brien}, {Obuchi}, {O'Dell}, {Ogaki}, {Oganesyan}, {Oh}, {Oh},
  {Oh}, {Ohashi}, {Ohishi}, {Ohkawa}, {Ohme}, {Ohta}, {Okada}, {Okutani}, {Okutomi}, {Olivetto}, {Oohara}, {Ooi}, {Oram}, {O'Reilly}, {Ormiston}, {Ormsby}, {Ortega}, {O'Shaughnessy}, {O'Shea}, {Oshino}, {Ossokine}, {Osthelder}, {Otabe}, {Ottaway}, {Overmier}, {Pace}, {Pagano}, {Page}, {Pagliaroli}, {Pai}, {Pai}, {Palamos}, {Palashov}, {Palomba}, {Pan}, {Panda}, {Pang}, {Pang}, {Pankow}, {Pannarale}, {Pant}, {Paoletti}, {Paoli}, {Paolone}, {Parisi}, {Park}, {Parker}, {Pascucci}, {Pasqualetti}, {Passaquieti}, {Passuello}, {Patel}, {Patricelli}, {Payne}, {Pechsiri}, {Pedraza}, {Pegoraro}, {Pele}, {Pe{\~n}a Arellano}, {Penn}, {Perego}, {Pereira}, {Pereira}, {Perez}, {P{\'e}rigois}, {Perreca}, {Perri{\`e}s}, {Petermann}, {Petterson}, {Pfeiffer}, {Pham}, {Phukon}, {Piccinni}, {Pichot}, {Piendibene}, {Piergiovanni}, {Pierini}, {Pierro}, {Pillant}, {Pilo}, {Pinard}, {Pinto}, {Piotrzkowski}, {Piotrzkowski}, {Pirello}, {Pitkin}, {Placidi}, {Plastino}, {Pluchar}, {Poggiani}, {Polini}, {Pong}, {Ponrathnam}, {Popolizio},
  {Porter}, {Powell}, {Pracchia}, {Pradier}, {Prajapati}, {Prasai}, {Prasanna}, {Pratten}, {Prestegard}, {Principe}, {Prodi}, {Prokhorov}, {Prosposito}, {Prudenzi}, {Puecher}, {Punturo}, {Puosi}, {Puppo}, {P{\"u}rrer}, {Qi}, {Quetschke}, {Quinonez}, {Quitzow-James}, {Raab}, {Raaijmakers}, {Radkins}, {Radulesco}, {Raffai}, {Rail}, {Raja}, {Rajan}, {Ramirez}, {Ramirez}, {Ramos-Buades}, {Rana}, {Rapagnani}, {Rapol}, {Ratto}, {Ray}, {Raymond}, {Raza}, {Razzano}, {Read}, {Rees}, {Regimbau}, {Rei}, {Reid}, {Reitze}, {Relton}, {Rettegno}, {Ricci}, {Richardson}, {Richardson}, {Richardson}, {Ricker}, {Riemenschneider}, {Riles}, {Rizzo}, {Robertson}, {Robie}, {Robinet}, {Rocchi}, {Rocha}, {Rodriguez}, {Rodriguez-Soto}, {Rolland}, {Rollins}, {Roma}, {Romanelli}, {Romano}, {Romel}, {Romero}, {Romero-Shaw}, {Romie}, {Rose}, {Rosi{\'n}ska}, {Rosofsky}, {Ross}, {Rowan}, {Rowlinson}, {Roy}, {Roy}, {Rozza}, {Ruggi}, {Ryan}, {Sachdev}, {Sadecki}, {Sadiq}, {Sago}, {Saito}, {Saito}, {Sakai}, {Sakai}, {Sakellariadou}, {Sakuno},
  {Salafia}, {Salconi}, {Saleem}, {Salemi}, {Samajdar}, {Sanchez}, {Sanchez}, {Sanchez}, {Sanchis-Gual}, {Sanders}, {Sanuy}, {Saravanan}, {Sarin}, {Sassolas}, {Satari}, {Sathyaprakash}, {Sato}, {Sato}, {Sauter}, {Savage}, {Savant}, {Sawada}, {Sawant}, {Sawant}, {Sayah}, {Schaetzl}, {Scheel}, {Scheuer}, {Schindler-Tyka}, {Schmidt}, {Schnabel}, {Schneewind}, {Schofield}, {Sch{\"o}nbeck}, {Schulte}, {Schutz}, {Schwartz}, {Scott}, {Scott}, {Seglar-Arroyo}, {Seidel}, {Sekiguchi}, {Sekiguchi}, {Sellers}, {Sengupta}, {Sennett}, {Sentenac}, {Seo}, {Sequino}, {Sergeev}, {Setyawati}, {Shaffer}, {Shahriar}, {Shams}, {Shao}, {Sharifi}, {Sharma}, {Sharma}, {Shawhan}, {Shcheblanov}, {Shen}, {Shibagaki}, {Shikauchi}, {Shimizu}, {Shimoda}, {Shimode}, {Shink}, {Shinkai}, {Shishido}, {Shoda}, {Shoemaker}, {Shoemaker}, {Shukla}, {Shyamsundar}, {Sieniawska}, {Sigg}, {Singer}, {Singh}, {Singh}, {Singha}, {Sintes}, {Sipala}, {Skliris}, {Slagmolen}, {Slaven-Blair}, {Smetana}, {Smith}, {Smith}, {Somala}, {Somiya}, {Son}, {Soni},
  {Soni}, {Sorazu}, {Sordini}, {Sorrentino}, {Sorrentino}, {Sotani}, {Soulard}, {Souradeep}, {Sowell}, {Spagnuolo}, {Spencer}, {Spera}, {Srivastava}, {Srivastava}, {Staats}, {Stachie}, {Steer}, {Steinlechner}, {Steinlechner}, {Stops}, {Stevenson}, {Stover}, {Strain}, {Strang}, {Stratta}, {Strunk}, {Sturani}, {Stuver}, {S{\"u}dbeck}, {Sudhagar}, {Sudhir}, {Sugimoto}, {Suh}, {Summerscales}, {Sun}, {Sun}, {Sunil}, {Sur}, {Suresh}, {Sutton}, {Suzuki}, {Suzuki}, {Swinkels}, {Szczepa{\'n}czyk}, {Szewczyk}, {Tacca}, {Tagoshi}, {Tait}, {Takahashi}, {Takahashi}, {Takamori}, {Takano}, {Takeda}, {Takeda}, {Talbot}, {Tanaka}, {Tanaka}, {Tanaka}, {Tanaka}, {Tanaka}, {Tanasijczuk}, {Tanioka}, {Tanner}, {Tao}, {Tapia}, {Tapia San Martin}, {Tasson}, {Telada}, {Tenorio}, {Terkowski}, {Test}, {Thirugnanasambandam}, {Thomas}, {Thomas}, {Thompson}, {Thondapu}, {Thorne}, {Thrane}, {Tiwari}, {Tiwari}, {Tiwari}, {Toland}, {Tolley}, {Tomaru}, {Tomigami}, {Tomura}, {Tonelli}, {Torres-Forn{\'e}}, {Torrie}, {Tosta E Melo},
  {T{\"o}yr{\"a}}, {Trapananti}, {Travasso}, {Traylor}, {Tringali}, {Tripathee}, {Troiano}, {Trovato}, {Trozzo}, {Trudeau}, {Tsai}, {Tsai}, {Tsang}, {Tsang}, {Tsao}, {Tse}, {Tso}, {Tsubono}, {Tsuchida}, {Tsukada}, {Tsuna}, {Tsutsui}, {Tsuzuki}, {Turconi}, {Tuyenbayev}, {Ubhi}, {Uchikata}, {Uchiyama}, {Udall}, {Ueda}, {Uehara}, {Ueno}, {Ueshima}, {Ugolini}, {Unnikrishnan}, {Uraguchi}, {Urban}, {Ushiba}, {Usman}, {Utina}, {Vahlbruch}, {Vajente}, {Vajpeyi}, {Valdes}, {Valentini}, {Valsan}, {van Bakel}, {van Beuzekom}, {van den Brand}, {van den Broeck}, {Vander-Hyde}, {van der Schaaf}, {van Heijningen}, {Vanosky}, {van Putten}, {Vardaro}, {Vargas}, {Varma}, {Vas{\'u}th}, {Vecchio}, {Vedovato}, {Veitch}, {Veitch}, {Venkateswara}, {Venneberg}, {Venugopalan}, {Verkindt}, {Verma}, {Veske}, {Vetrano}, {Vicer{\'e}}, {Viets}, {Villa-Ortega}, {Vinet}, {Vitale}, {Vo}, {Vocca}, {von Reis}, {von Wrangel}, {Vorvick}, {Vyatchanin}, {Wade}, {Wade}, {Wagner}, {Walet}, {Walker}, {Wallace}, {Wallace}, {Walsh}, {Wang}, {Wang},
  {Wang}, {Ward}, {Warner}, {Was}, {Washimi}, {Washington}, {Watchi}, {Weaver}, {Wei}, {Weinert}, {Weinstein}, {Weiss}, {Weller}, {Wellmann}, {Wen}, {We{\ss}els}, {Westhouse}, {Wette}, {Whelan}, {White}, {Whiting}, {Whittle}, {Wilken}, {Williams}, {Williams}, {Williamson}, {Willis}, {Willke}, {Wilson}, {Winkler}, {Wipf}, {Wlodarczyk}, {Woan}, {Woehler}, {Wofford}, {Wong}, {Wu}, {Wu}, {Wu}, {Wu}, {Wysocki}, {Xiao}, {Xu}, {Yamada}, {Yamamoto}, {Yamamoto}, {Yamamoto}, {Yamamoto}, {Yamashita}, {Yamazaki}, {Yang}, {Yang}, {Yang}, {Yang}, {Yang}, {Yap}, {Yeeles}, {Yelikar}, {Ying}, {Yokogawa}, {Yokoyama}, {Yokozawa}, {Yoon}, {Yoshioka}, {Yu}, {Yu}, {Yuzurihara}, {Zadro{\.z}ny}, {Zanolin}, {Zappa}, {Zeidler}, {Zelenova}, {Zendri}, {Zevin}, {Zhan}, {Zhang}, {Zhang}, {Zhang}, {Zhang}, {Zhang}, {Zhao}, {Zhao}, {Zhao}, {Zhao}, {Zhou}, {Zhu}, {Zhu}, {Zimmerman}, {Zlochower}, {Zucker}, {Zweizig}, {Ligo Scientific Collaboration}, {VIRGO Collaboration}, \& {KAGRA Collaboration}}]{Abbott2021BHNS}
{Abbott}, R., {Abbott}, T.~D., {Abraham}, S., {et~al.} 2021, \apjl, 915, L5

\bibitem[{{Acernese} {et~al.}(2015){Acernese}, {Agathos}, {Agatsuma}, {Aisa}, {Allemandou}, {Allocca}, {Amarni}, {Astone}, {Balestri}, {Ballardin}, {Barone}, {Baronick}, {Barsuglia}, {Basti}, {Basti}, {Bauer}, {Bavigadda}, {Bejger}, {Beker}, {Belczynski}, {Bersanetti}, {Bertolini}, {Bitossi}, {Bizouard}, {Bloemen}, {Blom}, {Boer}, {Bogaert}, {Bondi}, {Bondu}, {Bonelli}, {Bonnand}, {Boschi}, {Bosi}, {Bouedo}, {Bradaschia}, {Branchesi}, {Briant}, {Brillet}, {Brisson}, {Bulik}, {Bulten}, {Buskulic}, {Buy}, {Cagnoli}, {Calloni}, {Campeggi}, {Canuel}, {Carbognani}, {Cavalier}, {Cavalieri}, {Cella}, {Cesarini}, {Chassande-Mottin}, {Chincarini}, {Chiummo}, {Chua}, {Cleva}, {Coccia}, {Cohadon}, {Colla}, {Colombini}, {Conte}, {Coulon}, {Cuoco}, {Dalmaz}, {D'Antonio}, {Dattilo}, {Davier}, {Day}, {Debreczeni}, {Degallaix}, {Del{\'e}glise}, {Del Pozzo}, {Dereli}, {De Rosa}, {Di Fiore}, {Di Lieto}, {Di Virgilio}, {Doets}, {Dolique}, {Drago}, {Ducrot}, {Endr{\H{o}}czi}, {Fafone}, {Farinon}, {Ferrante}, {Ferrini},
  {Fidecaro}, {Fiori}, {Flaminio}, {Fournier}, {Franco}, {Frasca}, {Frasconi}, {Gammaitoni}, {Garufi}, {Gaspard}, {Gatto}, {Gemme}, {Gendre}, {Genin}, {Gennai}, {Ghosh}, {Giacobone}, {Giazotto}, {Gouaty}, {Granata}, {Greco}, {Groot}, {Guidi}, {Harms}, {Heidmann}, {Heitmann}, {Hello}, {Hemming}, {Hennes}, {Hofman}, {Jaranowski}, {Jonker}, {Kasprzack}, {K{\'e}f{\'e}lian}, {Kowalska}, {Kraan}, {Kr{\'o}lak}, {Kutynia}, {Lazzaro}, {Leonardi}, {Leroy}, {Letendre}, {Li}, {Lieunard}, {Lorenzini}, {Loriette}, {Losurdo}, {Magazz{\`u}}, {Majorana}, {Maksimovic}, {Malvezzi}, {Man}, {Mangano}, {Mantovani}, {Marchesoni}, {Marion}, {Marque}, {Martelli}, {Martellini}, {Masserot}, {Meacher}, {Meidam}, {Mezzani}, {Michel}, {Milano}, {Minenkov}, {Moggi}, {Mohan}, {Montani}, {Morgado}, {Mours}, {Mul}, {Nagy}, {Nardecchia}, {Naticchioni}, {Nelemans}, {Neri}, {Neri}, {Nocera}, {Pacaud}, {Palomba}, {Paoletti}, {Paoli}, {Pasqualetti}, {Passaquieti}, {Passuello}, {Perciballi}, {Petit}, {Pichot}, {Piergiovanni}, {Pillant}, {Piluso},
  {Pinard}, {Poggiani}, {Prijatelj}, {Prodi}, {Punturo}, {Puppo}, {Rabeling}, {R{\'a}cz}, {Rapagnani}, {Razzano}, {Re}, {Regimbau}, {Ricci}, {Robinet}, {Rocchi}, {Rolland}, {Romano}, {Rosi{\'n}ska}, {Ruggi}, {Saracco}, {Sassolas}, {Schimmel}, {Sentenac}, {Sequino}, {Shah}, {Siellez}, {Straniero}, {Swinkels}, {Tacca}, {Tonelli}, {Travasso}, {Turconi}, {Vajente}, {van Bakel}, {van Beuzekom}, {van den Brand}, {Van Den Broeck}, {van der Sluys}, {van Heijningen}, {Vas{\'u}th}, {Vedovato}, {Veitch}, {Verkindt}, {Vetrano}, {Vicer{\'e}}, {Vinet}, {Visser}, {Vocca}, {Ward}, {Was}, {Wei}, {Yvert}, {Zadro {\.z}ny}, \& {Zendri}}]{acernese2015}
{Acernese}, F., {Agathos}, M., {Agatsuma}, K., {et~al.} 2015, Classical and Quantum Gravity, 32, 024001

\bibitem[{{Ai} {et~al.}(2020){Ai}, {Gao}, \& {Zhang}}]{Ai2020}
{Ai}, S., {Gao}, H., \& {Zhang}, B. 2020, \apj, 893, 146

\bibitem[{{Aso} {et~al.}(2013){Aso}, {Michimura}, {Somiya}, {Ando}, {Miyakawa}, {Sekiguchi}, {Tatsumi}, \& {Yamamoto}}]{aso2013}
{Aso}, Y., {Michimura}, Y., {Somiya}, K., {et~al.} 2013, \prd, 88, 043007

\bibitem[{{Asplund} {et~al.}(2009){Asplund}, {Grevesse}, {Sauval}, \& {Scott}}]{Asplund2009}
{Asplund}, M., {Grevesse}, N., {Sauval}, A.~J., \& {Scott}, P. 2009, \araa, 47, 481

\bibitem[{{Bailyn} {et~al.}(1998){Bailyn}, {Jain}, {Coppi}, \& {Orosz}}]{Bailyn1998}
{Bailyn}, C.~D., {Jain}, R.~K., {Coppi}, P., \& {Orosz}, J.~A. 1998, \apj, 499, 367

\bibitem[{Bardeen(1970)}]{Bardeen1970}
Bardeen, J.~M. 1970, Nature, 226, 64

\bibitem[{{Bardeen} {et~al.}(1972){Bardeen}, {Press}, \& {Teukolsky}}]{Bardeen1972}
{Bardeen}, J.~M., {Press}, W.~H., \& {Teukolsky}, S.~A. 1972, \apj, 178, 347

\bibitem[{{Bavera} {et~al.}(2020){Bavera}, {Fragos}, {Qin}, {Zapartas}, {Neijssel}, {Mandel}, {Batta}, {Gaebel}, {Kimball}, \& {Stevenson}}]{Bavera2020}
{Bavera}, S.~S., {Fragos}, T., {Qin}, Y., {et~al.} 2020, \aap, 635, A97

\bibitem[{{Bavera} {et~al.}(2021){Bavera}, {Fragos}, {Zevin}, {Berry}, {Marchant}, {Andrews}, {Coughlin}, {Dotter}, {Kovlakas}, {Misra}, {Serra-Perez}, {Qin}, {Rocha}, {Rom{\'a}n-Garza}, {Tran}, \& {Zapartas}}]{Bavera2021}
{Bavera}, S.~S., {Fragos}, T., {Zevin}, M., {et~al.} 2021, \aap, 647, A153

\bibitem[{{Belczynski} {et~al.}(2020){Belczynski}, {Klencki}, {Fields}, {Olejak}, {Berti}, {Meynet}, {Fryer}, {Holz}, {O'Shaughnessy}, {Brown}, {Bulik}, {Leung}, {Nomoto}, {Madau}, {Hirschi}, {Kaiser}, {Jones}, {Mondal}, {Chruslinska}, {Drozda}, {Gerosa}, {Doctor}, {Giersz}, {Ekstrom}, {Georgy}, {Askar}, {Baibhav}, {Wysocki}, {Natan}, {Farr}, {Wiktorowicz}, {Coleman Miller}, {Farr}, \& {Lasota}}]{Belczynski2020}
{Belczynski}, K., {Klencki}, J., {Fields}, C.~E., {et~al.} 2020, \aap, 636, A104

\bibitem[{{Belczynski} {et~al.}(2012){Belczynski}, {Wiktorowicz}, {Fryer}, {Holz}, \& {Kalogera}}]{Belczynski2012}
{Belczynski}, K., {Wiktorowicz}, G., {Fryer}, C.~L., {Holz}, D.~E., \& {Kalogera}, V. 2012, \apj, 757, 91

\bibitem[{{Blandford} \& {Znajek}(1977)}]{Blandford1977}
{Blandford}, R.~D. \& {Znajek}, R.~L. 1977, \mnras, 179, 433

\bibitem[{{B{\"o}hm-Vitense}(1958)}]{MLT1958}
{B{\"o}hm-Vitense}, E. 1958, \zap, 46, 108

\bibitem[{{Briel} {et~al.}(2023){Briel}, {Stevance}, \& {Eldridge}}]{Briel2023}
{Briel}, M.~M., {Stevance}, H.~F., \& {Eldridge}, J.~J. 2023, \mnras, 520, 5724

\bibitem[{{Broekgaarden} {et~al.}(2021){Broekgaarden}, {Berger}, {Neijssel}, {Vigna-G{\'o}mez}, {Chattopadhyay}, {Stevenson}, {Chruslinska}, {Justham}, {de Mink}, \& {Mandel}}]{Broekgaarden2021}
{Broekgaarden}, F.~S., {Berger}, E., {Neijssel}, C.~J., {et~al.} 2021, \mnras, 508, 5028

\bibitem[{{Chaboyer} \& {Zahn}(1992)}]{Chaboyer1992}
{Chaboyer}, B. \& {Zahn}, J.~P. 1992, \aap, 253, 173

\bibitem[{{Chandra} {et~al.}(2024){Chandra}, {Gupta}, {Gamba}, {Kashyap}, {Chattopadhyay}, {Gonzalez}, {Bernuzzi}, \& {Sathyaprakash}}]{Chandra2024}
{Chandra}, K., {Gupta}, I., {Gamba}, R., {et~al.} 2024, arXiv e-prints, arXiv:2405.03841

\bibitem[{{Chattopadhyay} {et~al.}(2021){Chattopadhyay}, {Stevenson}, {Hurley}, {Bailes}, \& {Broekgaarden}}]{Chattopadhyay2021}
{Chattopadhyay}, D., {Stevenson}, S., {Hurley}, J.~R., {Bailes}, M., \& {Broekgaarden}, F. 2021, \mnras, 504, 3682

\bibitem[{{de Mink} {et~al.}(2013){de Mink}, {Langer}, {Izzard}, {Sana}, \& {de Koter}}]{Demink2013}
{de Mink}, S.~E., {Langer}, N., {Izzard}, R.~G., {Sana}, H., \& {de Koter}, A. 2013, \apj, 764, 166

\bibitem[{{Detmers} {et~al.}(2008){Detmers}, {Langer}, {Podsiadlowski}, \& {Izzard}}]{Detmers2008}
{Detmers}, R.~G., {Langer}, N., {Podsiadlowski}, P., \& {Izzard}, R.~G. 2008, \aap, 484, 831

\bibitem[{{Fan} {et~al.}(2024){Fan}, {Han}, {Jiang}, {Shao}, \& {Tang}}]{Fan2024}
{Fan}, Y.-Z., {Han}, M.-Z., {Jiang}, J.-L., {Shao}, D.-S., \& {Tang}, S.-P. 2024, \prd, 109, 043052

\bibitem[{{Farr} {et~al.}(2011){Farr}, {Sravan}, {Cantrell}, {Kreidberg}, {Bailyn}, {Mandel}, \& {Kalogera}}]{Farr2011}
{Farr}, W.~M., {Sravan}, N., {Cantrell}, A., {et~al.} 2011, \apj, 741, 103

\bibitem[{{Foucart} {et~al.}(2018){Foucart}, {Hinderer}, \& {Nissanke}}]{Foucart2018}
{Foucart}, F., {Hinderer}, T., \& {Nissanke}, S. 2018, \prd, 98, 081501

\bibitem[{{Fragos} {et~al.}(2023){Fragos}, {Andrews}, {Bavera}, {Berry}, {Coughlin}, {Dotter}, {Giri}, {Kalogera}, {Katsaggelos}, {Kovlakas}, {Lalvani}, {Misra}, {Srivastava}, {Qin}, {Rocha}, {Rom{\'a}n-Garza}, {Serra}, {Stahle}, {Sun}, {Teng}, {Trajcevski}, {Tran}, {Xing}, {Zapartas}, \& {Zevin}}]{Tassos2023}
{Fragos}, T., {Andrews}, J.~J., {Bavera}, S.~S., {et~al.} 2023, \apjs, 264, 45

\bibitem[{{Fragos} \& {McClintock}(2015)}]{Fragos2015}
{Fragos}, T. \& {McClintock}, J.~E. 2015, \apj, 800, 17

\bibitem[{{Fryer} {et~al.}(2012){Fryer}, {Belczynski}, {Wiktorowicz}, {Dominik}, {Kalogera}, \& {Holz}}]{Fryer2012}
{Fryer}, C.~L., {Belczynski}, K., {Wiktorowicz}, G., {et~al.} 2012, \apj, 749, 91

\bibitem[{{Fuller} \& {Ma}(2019)}]{Fuller2019}
{Fuller}, J. \& {Ma}, L. 2019, \apjl, 881, L1

\bibitem[{{Gallegos-Garcia} {et~al.}(2023){Gallegos-Garcia}, {Berry}, \& {Kalogera}}]{Monica2023}
{Gallegos-Garcia}, M., {Berry}, C. P.~L., \& {Kalogera}, V. 2023, \apj, 955, 133

\bibitem[{{Gallegos-Garcia} {et~al.}(2021){Gallegos-Garcia}, {Berry}, {Marchant}, \& {Kalogera}}]{Monica2021}
{Gallegos-Garcia}, M., {Berry}, C. P.~L., {Marchant}, P., \& {Kalogera}, V. 2021, \apj, 922, 110

\bibitem[{{Heger} \& {Langer}(1998)}]{Heger1998}
{Heger}, A. \& {Langer}, N. 1998, \aap, 334, 210

\bibitem[{{Heger} \& {Langer}(2000)}]{Heger2000}
{Heger}, A. \& {Langer}, N. 2000, \apj, 544, 1016

\bibitem[{{Hu} {et~al.}(2023){Hu}, {Zhu}, {Qin}, {Shao}, {Zhang}, {Yu}, {Liang}, {Liu}, {Wang}, {Shu}, \& {Liu}}]{Hu2023}
{Hu}, R.-C., {Zhu}, J.-P., {Qin}, Y., {et~al.} 2023, arXiv e-prints, arXiv:2301.06402

\bibitem[{{Hu} {et~al.}(2022){Hu}, {Zhu}, {Qin}, {Zhang}, {Liang}, \& {Shao}}]{Hu2022}
{Hu}, R.-C., {Zhu}, J.-P., {Qin}, Y., {et~al.} 2022, \apj, 928, 163

\bibitem[{{Hunter}(2007)}]{Hunter2007}
{Hunter}, J.~D. 2007, Computing in Science and Engineering, 9, 90

\bibitem[{{Hurley} {et~al.}(2002){Hurley}, {Tout}, \& {Pols}}]{Hurley2002}
{Hurley}, J.~R., {Tout}, C.~A., \& {Pols}, O.~R. 2002, \mnras, 329, 897

\bibitem[{{Hut}(1981)}]{Hut1981}
{Hut}, P. 1981, \aap, 99, 126

\bibitem[{{Jermyn} {et~al.}(2023){Jermyn}, {Bauer}, {Schwab}, {Farmer}, {Ball}, {Bellinger}, {Dotter}, {Joyce}, {Marchant}, {Mombarg}, {Wolf}, {Sunny Wong}, {Cinquegrana}, {Farrell}, {Smolec}, {Thoul}, {Cantiello}, {Herwig}, {Toloza}, {Bildsten}, {Townsend}, \& {Timmes}}]{Paxton2023}
{Jermyn}, A.~S., {Bauer}, E.~B., {Schwab}, J., {et~al.} 2023, \apjs, 265, 15

\bibitem[{{Kalogera} \& {Baym}(1996)}]{Kalogera1996}
{Kalogera}, V. \& {Baym}, G. 1996, \apjl, 470, L61

\bibitem[{{Kippenhahn} \& {Weigert}(1990)}]{Kippenhahn1990}
{Kippenhahn}, R. \& {Weigert}, A. 1990, {Stellar Structure and Evolution}

\bibitem[{{Kolb} \& {Ritter}(1990)}]{Kolb1990}
{Kolb}, U. \& {Ritter}, H. 1990, \aap, 236, 385

\bibitem[{{Kunnumkai} {et~al.}(2024){Kunnumkai}, {Palmese}, {Bulla}, {Dietrich}, {Farah}, \& {Pang}}]{Kunnumkai2024}
{Kunnumkai}, K., {Palmese}, A., {Bulla}, M., {et~al.} 2024, arXiv e-prints, arXiv:2409.10651

\bibitem[{{Kyutoku} {et~al.}(2015){Kyutoku}, {Ioka}, {Okawa}, {Shibata}, \& {Taniguchi}}]{Kyutoku2015}
{Kyutoku}, K., {Ioka}, K., {Okawa}, H., {Shibata}, M., \& {Taniguchi}, K. 2015, \prd, 92, 044028

\bibitem[{{Langer}(1998)}]{Langer1998}
{Langer}, N. 1998, \aap, 329, 551

\bibitem[{{Langer} {et~al.}(1983){Langer}, {Fricke}, \& {Sugimoto}}]{Langer1983}
{Langer}, N., {Fricke}, K.~J., \& {Sugimoto}, D. 1983, \aap, 126, 207

\bibitem[{{Lattimer} \& {Yahil}(1989)}]{Lattimer1989}
{Lattimer}, J.~M. \& {Yahil}, A. 1989, \apj, 340, 426

\bibitem[{{Lei} {et~al.}(2017){Lei}, {Zhang}, {Wu}, \& {Liang}}]{Lei2017}
{Lei}, W.-H., {Zhang}, B., {Wu}, X.-F., \& {Liang}, E.-W. 2017, \apj, 849, 47

\bibitem[{{Li} \& {Paczy{\'n}ski}(1998)}]{Li1998}
{Li}, L.-X. \& {Paczy{\'n}ski}, B. 1998, \apjl, 507, L59

\bibitem[{{Lyu} {et~al.}(2023){Lyu}, {Yuan}, {Wu}, {Guo}, {Wang}, {Yi}, {Tang}, {Hu}, {Zhu}, {Shu}, {Qin}, \& {Liang}}]{lv2023}
{Lyu}, F., {Yuan}, L., {Wu}, D.~H., {et~al.} 2023, \mnras, 525, 4321

\bibitem[{{Ma} \& {Fuller}(2023)}]{Ma2023}
{Ma}, L. \& {Fuller}, J. 2023, \apj, 952, 53

\bibitem[{{Maeder} \& {Meynet}(2000)}]{Maeder2000}
{Maeder}, A. \& {Meynet}, G. 2000, \aap, 361, 159

\bibitem[{{Marchant} {et~al.}(2017){Marchant}, {Langer}, {Podsiadlowski}, {Tauris}, {de Mink}, {Mandel}, \& {Moriya}}]{Marchant2017}
{Marchant}, P., {Langer}, N., {Podsiadlowski}, P., {et~al.} 2017, \aap, 604, A55

\bibitem[{{Margalit} \& {Metzger}(2017)}]{Margalit2017}
{Margalit}, B. \& {Metzger}, B.~D. 2017, \apjl, 850, L19

\bibitem[{{Metzger} {et~al.}(2010){Metzger}, {Mart{\'\i}nez-Pinedo}, {Darbha}, {Quataert}, {Arcones}, {Kasen}, {Thomas}, {Nugent}, {Panov}, \& {Zinner}}]{Metzger2010}
{Metzger}, B.~D., {Mart{\'\i}nez-Pinedo}, G., {Darbha}, S., {et~al.} 2010, \mnras, 406, 2650

\bibitem[{{Motta} {et~al.}(2017){Motta}, {Kajava}, {S{\'a}nchez-Fern{\'a}ndez}, {Beardmore}, {Sanna}, {Page}, {Fender}, {Altamirano}, {Charles}, {Giustini}, {Knigge}, {Kuulkers}, {Oates}, \& {Osborne}}]{Motta2017}
{Motta}, S.~E., {Kajava}, J.~J.~E., {S{\'a}nchez-Fern{\'a}ndez}, C., {et~al.} 2017, \mnras, 471, 1797

\bibitem[{{M{\"u}ller} \& {Serot}(1996)}]{Muller1996}
{M{\"u}ller}, H. \& {Serot}, B.~D. 1996, \nphysa, 606, 508

\bibitem[{{Narayan} {et~al.}(1992){Narayan}, {Paczynski}, \& {Piran}}]{Narayan1992}
{Narayan}, R., {Paczynski}, B., \& {Piran}, T. 1992, \apjl, 395, L83

\bibitem[{{Neijssel} {et~al.}(2019){Neijssel}, {Vigna-G{\'o}mez}, {Stevenson}, {Barrett}, {Gaebel}, {Broekgaarden}, {de Mink}, {Sz{\'e}csi}, {Vinciguerra}, \& {Mandel}}]{Neijssel2019}
{Neijssel}, C.~J., {Vigna-G{\'o}mez}, A., {Stevenson}, S., {et~al.} 2019, \mnras, 490, 3740

\bibitem[{{Neilsen} {et~al.}(2016){Neilsen}, {Rahoui}, {Homan}, \& {Buxton}}]{Neilsen2016}
{Neilsen}, J., {Rahoui}, F., {Homan}, J., \& {Buxton}, M. 2016, \apj, 822, 20

\bibitem[{{Nitz} {et~al.}(2023){Nitz}, {Kumar}, {Wang}, {Kastha}, {Wu}, {Sch{\"a}fer}, {Dhurkunde}, \& {Capano}}]{nitz2021}
{Nitz}, A.~H., {Kumar}, S., {Wang}, Y.-F., {et~al.} 2023, \apj, 946, 59

\bibitem[{{Olejak} {et~al.}(2022){Olejak}, {Fryer}, {Belczynski}, \& {Baibhav}}]{Olejak2022}
{Olejak}, A., {Fryer}, C.~L., {Belczynski}, K., \& {Baibhav}, V. 2022, \mnras, 516, 2252

\bibitem[{{{\"O}zel} \& {Freire}(2016)}]{Ozel2016}
{{\"O}zel}, F. \& {Freire}, P. 2016, \araa, 54, 401

\bibitem[{{{\"O}zel} {et~al.}(2010){{\"O}zel}, {Psaltis}, {Narayan}, \& {McClintock}}]{Ozel2010}
{{\"O}zel}, F., {Psaltis}, D., {Narayan}, R., \& {McClintock}, J.~E. 2010, \apj, 725, 1918

\bibitem[{{Paczynski}(1991)}]{Paczynski1991}
{Paczynski}, B. 1991, \actaa, 41, 257

\bibitem[{{Paxton} {et~al.}(2011){Paxton}, {Bildsten}, {Dotter}, {Herwig}, {Lesaffre}, \& {Timmes}}]{Paxton2011}
{Paxton}, B., {Bildsten}, L., {Dotter}, A., {et~al.} 2011, \apjs, 192, 3

\bibitem[{{Paxton} {et~al.}(2013){Paxton}, {Cantiello}, {Arras}, {Bildsten}, {Brown}, {Dotter}, {Mankovich}, {Montgomery}, {Stello}, {Timmes}, \& {Townsend}}]{Paxton2013}
{Paxton}, B., {Cantiello}, M., {Arras}, P., {et~al.} 2013, \apjs, 208, 4

\bibitem[{{Paxton} {et~al.}(2015){Paxton}, {Marchant}, {Schwab}, {Bauer}, {Bildsten}, {Cantiello}, {Dessart}, {Farmer}, {Hu}, {Langer}, {Townsend}, {Townsley}, \& {Timmes}}]{Paxton2015}
{Paxton}, B., {Marchant}, P., {Schwab}, J., {et~al.} 2015, \apjs, 220, 15

\bibitem[{{Paxton} {et~al.}(2018){Paxton}, {Schwab}, {Bauer}, {Bildsten}, {Blinnikov}, {Duffell}, {Farmer}, {Goldberg}, {Marchant}, {Sorokina}, {Thoul}, {Townsend}, \& {Timmes}}]{Paxton2018}
{Paxton}, B., {Schwab}, J., {Bauer}, E.~B., {et~al.} 2018, \apjs, 234, 34

\bibitem[{{Paxton} {et~al.}(2019){Paxton}, {Smolec}, {Schwab}, {Gautschy}, {Bildsten}, {Cantiello}, {Dotter}, {Farmer}, {Goldberg}, {Jermyn}, {Kanbur}, {Marchant}, {Thoul}, {Townsend}, {Wolf}, {Zhang}, \& {Timmes}}]{Paxton2019}
{Paxton}, B., {Smolec}, R., {Schwab}, J., {et~al.} 2019, \apjs, 243, 10

\bibitem[{{Peters}(1964)}]{Peters1964}
{Peters}, P.~C. 1964, Physical Review, 136, 1224

\bibitem[{{Podsiadlowski} {et~al.}(2003){Podsiadlowski}, {Rappaport}, \& {Han}}]{Podsiadlowski2003}
{Podsiadlowski}, P., {Rappaport}, S., \& {Han}, Z. 2003, \mnras, 341, 385

\bibitem[{{Qin} {et~al.}(2018){Qin}, {Fragos}, {Meynet}, {Andrews}, {S{\o}rensen}, \& {Song}}]{Qin2018}
{Qin}, Y., {Fragos}, T., {Meynet}, G., {et~al.} 2018, \aap, 616, A28

\bibitem[{{Qin} {et~al.}(2023){Qin}, {Hu}, {Meynet}, {Wang}, {Zhu}, {Song}, {Shu}, \& {Wu}}]{Qin2023}
{Qin}, Y., {Hu}, R.~C., {Meynet}, G., {et~al.} 2023, \aap, 671, A62

\bibitem[{{Qin} {et~al.}(2019){Qin}, {Marchant}, {Fragos}, {Meynet}, \& {Kalogera}}]{Qin2019}
{Qin}, Y., {Marchant}, P., {Fragos}, T., {Meynet}, G., \& {Kalogera}, V. 2019, \apjl, 870, L18

\bibitem[{{Qin} {et~al.}(2022){Qin}, {Shu}, {Yi}, \& {Wang}}]{Qin2022}
{Qin}, Y., {Shu}, X., {Yi}, S., \& {Wang}, Y.-Z. 2022, Research in Astronomy and Astrophysics, 22, 035023

\bibitem[{{Rhoades} \& {Ruffini}(1974)}]{Rhoades1974}
{Rhoades}, C.~E. \& {Ruffini}, R. 1974, \prl, 32, 324

\bibitem[{{Ritter}(1988)}]{Ritter1988}
{Ritter}, H. 1988, \aap, 202, 93

\bibitem[{{Sciarini} {et~al.}(2024){Sciarini}, {Ekstr{\"o}m}, {Eggenberger}, {Meynet}, {Fragos}, \& {Song}}]{Sciarini2024}
{Sciarini}, L., {Ekstr{\"o}m}, S., {Eggenberger}, P., {et~al.} 2024, \aap, 681, L1

\bibitem[{{Shao} {et~al.}(2020){Shao}, {Tang}, {Jiang}, \& {Fan}}]{Shao2020}
{Shao}, D.-S., {Tang}, S.-P., {Jiang}, J.-L., \& {Fan}, Y.-Z. 2020, \prd, 102, 063006

\bibitem[{{Shao}(2022)}]{shao2022RAA}
{Shao}, Y. 2022, Research in Astronomy and Astrophysics, 22, 122002

\bibitem[{{Shao} \& {Li}(2022)}]{Shao2022}
{Shao}, Y. \& {Li}, X.-D. 2022, \apj, 930, 26

\bibitem[{{Somiya}(2012)}]{Somiya2012}
{Somiya}, K. 2012, Classical and Quantum Gravity, 29, 124007

\bibitem[{{Stevenson} {et~al.}(2017){Stevenson}, {Vigna-G{\'o}mez}, {Mandel}, {Barrett}, {Neijssel}, {Perkins}, \& {de Mink}}]{Stevenson2017}
{Stevenson}, S., {Vigna-G{\'o}mez}, A., {Mandel}, I., {et~al.} 2017, Nature Communications, 8, 14906

\bibitem[{{Team COMPAS: Riley, J.} {et~al.}(2022){Team COMPAS: Riley, J.}, {Agrawal}, {Barrett}, {Boyett}, {Broekgaarden}, {Chattopadhyay}, {Gaebel}, {Gittins}, {Hirai}, {Howitt}, {Justham}, {Khandelwal}, {Kummer}, {Lau}, {Mandel}, {de Mink}, {Neijssel}, {Riley}, {van Son}, {Stevenson}, {Vigna-Gomez}, {Vinciguerra}, {Wagg}, \& {Willcox}}]{TeamCOMPAS2022}
{Team COMPAS: Riley, J.}, {Agrawal}, P., {Barrett}, J.~W., {et~al.} 2022, \apjs, 258, 34

\bibitem[{{Timmes} {et~al.}(1996){Timmes}, {Woosley}, \& {Weaver}}]{Timmes1996}
{Timmes}, F.~X., {Woosley}, S.~E., \& {Weaver}, T.~A. 1996, \apj, 457, 834

\bibitem[{{Vigna-G{\'o}mez} {et~al.}(2018){Vigna-G{\'o}mez}, {Neijssel}, {Stevenson}, {Barrett}, {Belczynski}, {Justham}, {de Mink}, {M{\"u}ller}, {Podsiadlowski}, {Renzo}, {Sz{\'e}csi}, \& {Mandel}}]{Vigna2018}
{Vigna-G{\'o}mez}, A., {Neijssel}, C.~J., {Stevenson}, S., {et~al.} 2018, \mnras, 481, 4009

\bibitem[{{Vink} {et~al.}(2001){Vink}, {de Koter}, \& {Lamers}}]{vink2001}
{Vink}, J.~S., {de Koter}, A., \& {Lamers}, H.~J.~G.~L.~M. 2001, \aap, 369, 574

\bibitem[{{Wang} {et~al.}(2024{\natexlab{a}}){Wang}, {Zhao}, {Feng}, {Ge}, {Shao}, {Cui}, {Gao}, {Zhang}, {Wang}, {Li}, {Bai}, {Yuan}, {Huang}, {Yuan}, {Zhang}, {Yi}, {Xiang}, {Li}, {Li}, {Zhang}, {Zhang}, {Han}, {Fan}, {Li}, {Chen}, {Liu}, {Meng}, {Liu}, {Zhang}, {Gu}, \& {Liu}}]{Wangsong2024NA}
{Wang}, S., {Zhao}, X., {Feng}, F., {et~al.} 2024{\natexlab{a}}, Nature Astronomy [\eprint[arXiv]{2409.06352}]

\bibitem[{{Wang} \& {Zhang}(2024)}]{Wang2024NA}
{Wang}, Y. \& {Zhang}, B. 2024, Nature Astronomy [\eprint[arXiv]{2403.06416}]

\bibitem[{{Wang} {et~al.}(2024{\natexlab{b}}){Wang}, {Hu}, {Qin}, {Zhu}, {Zhang}, {Yi}, {Tang}, {Shu}, {Lyu}, \& {Liang}}]{Wang2024}
{Wang}, Z.-H.-T., {Hu}, R.-C., {Qin}, Y., {et~al.} 2024{\natexlab{b}}, \apj, 965, 177

\bibitem[{{Xing} {et~al.}(2024){Xing}, {Bavera}, {Fragos}, {Kruckow}, {Rom{\'a}n-Garza}, {Andrews}, {Dotter}, {Kovlakas}, {Misra}, {Srivastava}, {Rocha}, {Sun}, \& {Zapartas}}]{Xing2024}
{Xing}, Z., {Bavera}, S.~S., {Fragos}, T., {et~al.} 2024, \aap, 683, A144

\bibitem[{{Ye} {et~al.}(2024){Ye}, {Kremer}, {Ransom}, \& {Rasio}}]{Ye2024}
{Ye}, C.~S., {Kremer}, K., {Ransom}, S.~M., \& {Rasio}, F.~A. 2024, \apj, 975, 77

\bibitem[{{Zahn}(1977)}]{Zahn1977}
{Zahn}, J.~P. 1977, \aap, 57, 383

\bibitem[{{Zevin} \& {Bavera}(2022)}]{Zevin2022}
{Zevin}, M. \& {Bavera}, S.~S. 2022, \apj, 933, 86

\bibitem[{{Zevin} {et~al.}(2020){Zevin}, {Spera}, {Berry}, \& {Kalogera}}]{Zevin2020}
{Zevin}, M., {Spera}, M., {Berry}, C. P.~L., \& {Kalogera}, V. 2020, \apjl, 899, L1

\bibitem[{{Zhang}(2018)}]{Zhang2018}
{Zhang}, B. 2018, {The Physics of Gamma-Ray Bursts}

\bibitem[{{Zhang} {et~al.}(2023){Zhang}, {Wang}, {Zhu}, {Hu}, {Shu}, {Tang}, {Yi}, {Lyu}, {Liang}, \& {Qin}}]{Zhang2023}
{Zhang}, W.~T., {Wang}, Z.~H.~T., {Zhu}, J.~P., {et~al.} 2023, \mnras, 526, 854

\bibitem[{{Zhu} {et~al.}(2024{\natexlab{a}}){Zhu}, {Hu}, {Kang}, {Zhang}, {Tong}, {Shao}, \& {Qin}}]{Zhu2024GW230529}
{Zhu}, J.-P., {Hu}, R.-C., {Kang}, Y., {et~al.} 2024{\natexlab{a}}, \apj, 974, 211

\bibitem[{{Zhu} {et~al.}(2024{\natexlab{b}}){Zhu}, {Qin}, {Wang}, {Hu}, {Zhang}, \& {Wu}}]{Zhu2024SuperEdd}
{Zhu}, J.-P., {Qin}, Y., {Wang}, Z.-H.-T., {et~al.} 2024{\natexlab{b}}, \mnras, 529, 4554

\bibitem[{{Zhu} {et~al.}(2021{\natexlab{a}}){Zhu}, {Wu}, {Yang}, {Zhang}, {Gao}, {Yu}, {Li}, {Cao}, {Liu}, {Huang}, \& {Zhang}}]{Zhu2021a}
{Zhu}, J.-P., {Wu}, S., {Yang}, Y.-P., {et~al.} 2021{\natexlab{a}}, \apj, 917, 24

\bibitem[{{Zhu} {et~al.}(2021{\natexlab{b}}){Zhu}, {Wu}, {Yang}, {Zhang}, {Yu}, {Gao}, {Cao}, \& {Liu}}]{Zhu2021b}
{Zhu}, J.-P., {Wu}, S., {Yang}, Y.-P., {et~al.} 2021{\natexlab{b}}, \apj, 921, 156

\bibitem[{{Zhu} {et~al.}(2020){Zhu}, {Yang}, {Liu}, {Huang}, {Zhang}, {Li}, {Yu}, \& {Gao}}]{Zhu2020}
{Zhu}, J.-P., {Yang}, Y.-P., {Liu}, L.-D., {et~al.} 2020, \apj, 897, 20

\end{thebibliography}

\begin{appendix}
\section{}
\begin{figure*}[h]
     \centering
     \includegraphics[width=0.99\textwidth]{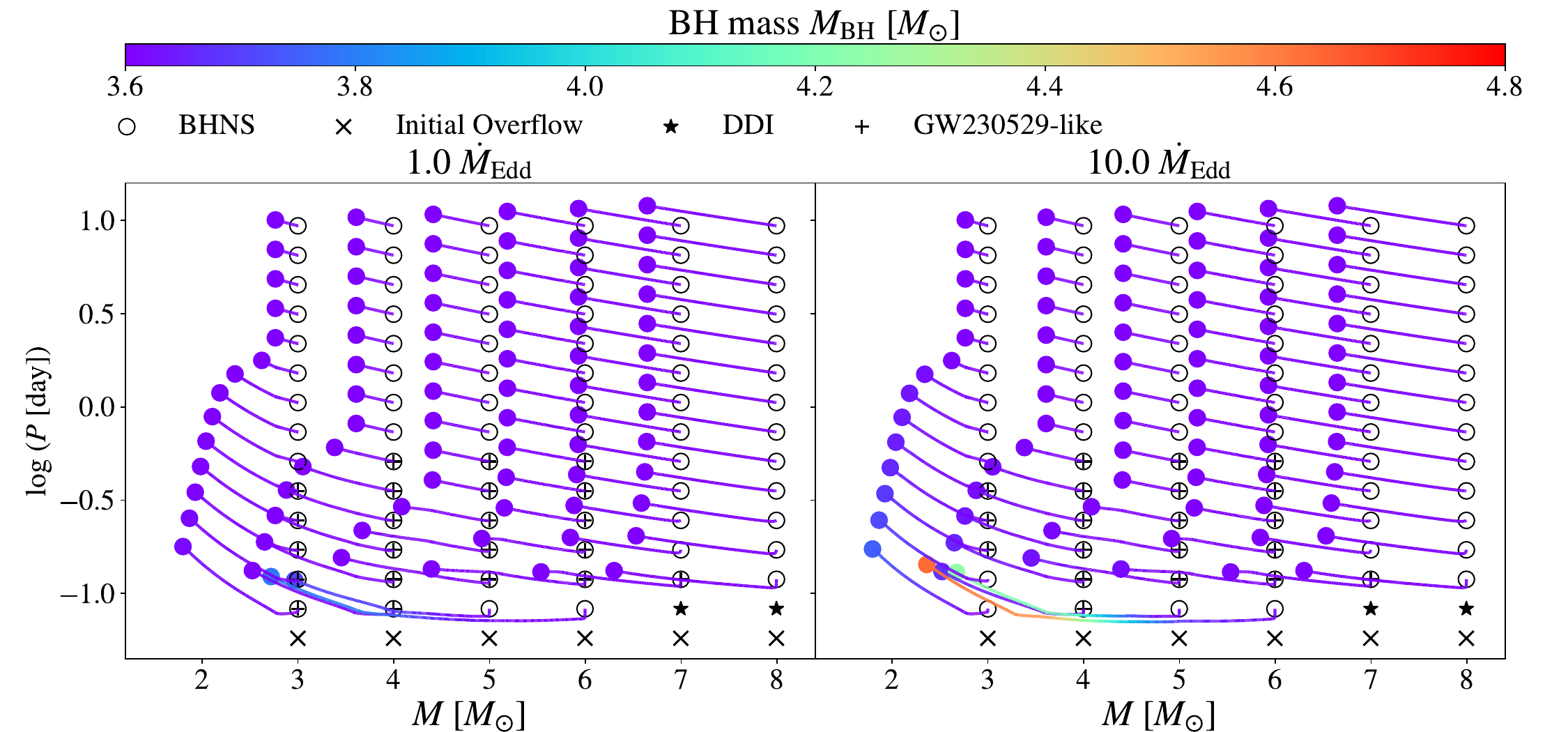}
     \caption{BH mass acquired through Roche-lobe accretion from a He-rich star, plotted against the star's mass and the orbital period. Black and colored symbols represent the initial and final values for the He-rich star mass and orbital period. The color gradient along the lines indicates the evolving mass of the BH as a function of the He-rich star mass and orbital period. Symbols denote different scenarios: circles for BHNS systems, stars for dynamical delayed instability (DDI), and crosses for initial overflow cases. The two panels correspond to different accretion rates: 1.0 $\dot{M}_{\rm Edd}$ in the \textit{left panel} and 10.0 $\dot{M}_{\rm Edd}$ in the \textit{right panel}. The plus symbol refers to the systems that might be representative of the progenitors of GW230529.}
     \label{fig5}
\end{figure*}

\begin{figure*}[h]
     \centering
     \includegraphics[width=0.99\textwidth]{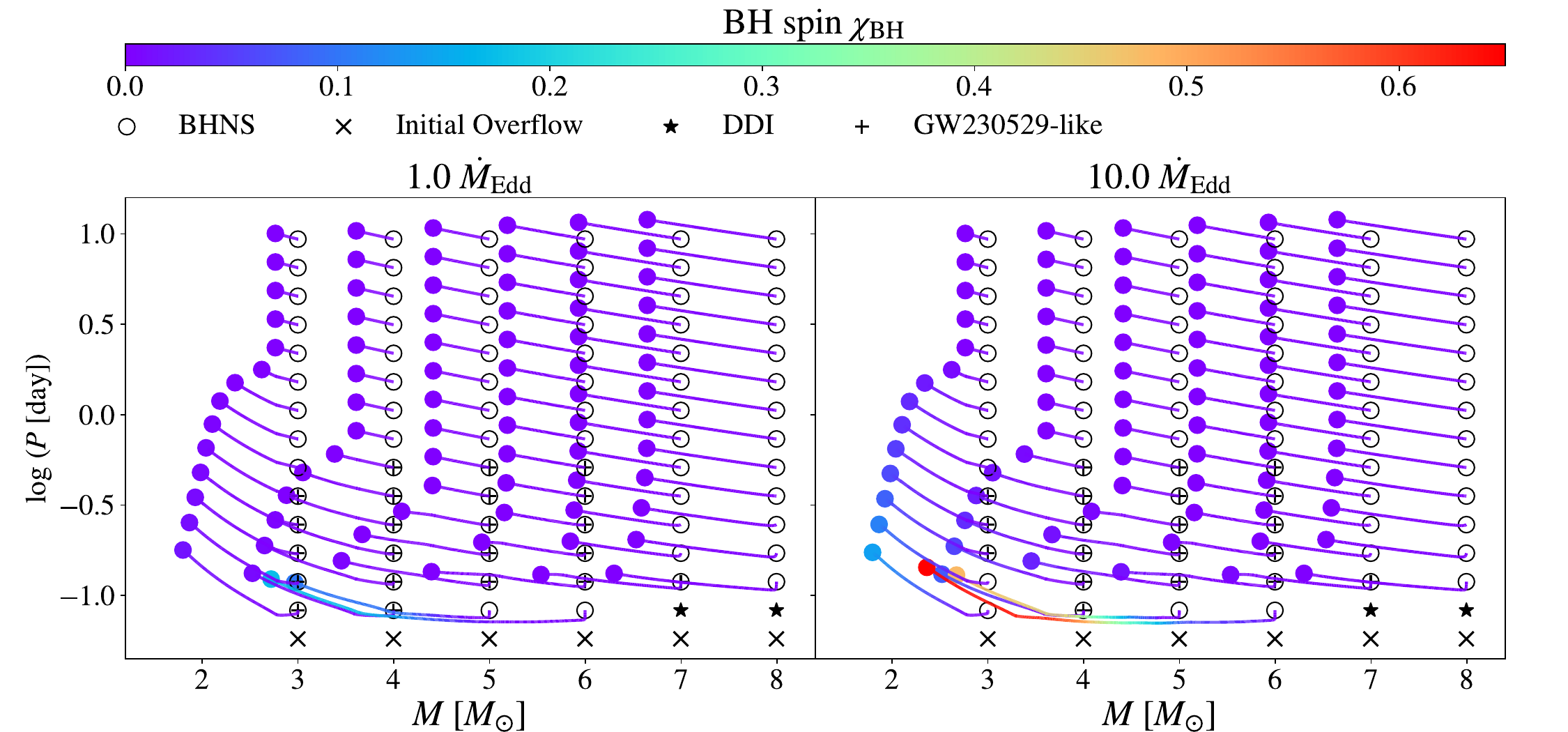}
     \caption{Similar to Figure \ref{fig5}, but the color representing the BH spin $\chi_{\rm BH}$.}
     \label{fig6}
\end{figure*}

The synchronization timescale ($\tau_{\rm sync}$) for dynamical tides proposed by \cite{Sciarini2024} is given as:

\begin{equation}
    \centering
    \begin{split}
    \frac{1}{\tau_{\rm sync}}\bigg|_{\rm Dyn}\equiv\frac{\rm d}{\rm dt}\bigg|\frac{\Omega_{\rm spin}-\Omega_{\rm orb}}{\Omega_{\rm orb}}\bigg|^{-5/3}\Bigg|_{e\approx0,\Omega_{\rm orb}={\rm const},I={\rm const}}= \\5\cdot2^{5/3}(\frac{GM}{R^3})^{1/2}\frac{MR^2}{I}q^2(1+q)^{5/6}E_2(\frac{R}{a})^{17/2},
     \end{split}
\end{equation}
where  $q = M_2/M$ is the mass ratio, $I$ the moment of inertia, $R$ the stellar radius, $a$ the semimajor axis, $E_2$ tidal torque coefficient. Assuming constant value for $\Omega_{\rm orb}$ and $\tau_{\rm sync}$, \cite{Sciarini2024} provided the analytical solution as follows.
\begin{equation}
    \frac{\Omega_{\rm spin}(t)-\Omega_{\rm orb}}{\Omega_{\rm orb}} = [(\frac{\Omega_{\rm 0}-\Omega_{\rm orb}}{\Omega_{\rm orb}})^{-5/3}+{\rm sgn}(\Omega_{\rm 0}-\Omega_{\rm orb})\frac{t}{\tau_{\rm sync}}]^{-3/5},
\end{equation}
where $\Omega_{\rm 0} = \Omega_{\rm spin}(t = 0)$. Assuming that $\tau_{{\rm sync}}$ and $\Omega_{\rm orb}$ are constant through a step \citep{Detmers2008,Paxton2015}, thus we obtain 

\begin{equation}
    \centering
    \begin{split}
    \Delta\Omega_{i.j}=\Omega_{\rm orb}\{[(\frac{\Omega_{i.j}-\Omega_{\rm orb}}{\Omega_{\rm orb}})^{-5/3}+{\rm sgn}(\Omega_{i,j,{\rm 0}}-\Omega_{\rm orb})\frac{\delta t}{\tau_{{\rm sync},j}}]^{-3/5}\\-\frac{\Omega_{i.j}-\Omega_{\rm orb}}{\Omega_{\rm orb}}\},
     \end{split}
\end{equation}
where $j$ = 1, 2 is the index of each star, $\Omega_{i.j}$ is the angular frequency at the face of cell $i$ toward the surface, and $\Omega_{i,j,{\rm 0}} = \Omega_{i.j}(t=0)$.
After implementing the above equation in the \texttt{MESA} binary module, we computed several models using different definitions of $\tau_{{\rm sync}}$. In these tests, we evolved
a binary system consisting of a BH and a He-rich star with identical initial masses but varying initial orbital periods (0.2, 0.5, and 1.0 days). We assumed that the He-rich star was initially synchronized with its orbit.

\begin{figure*}[h]
     \centering
     \includegraphics[width=0.99\columnwidth]{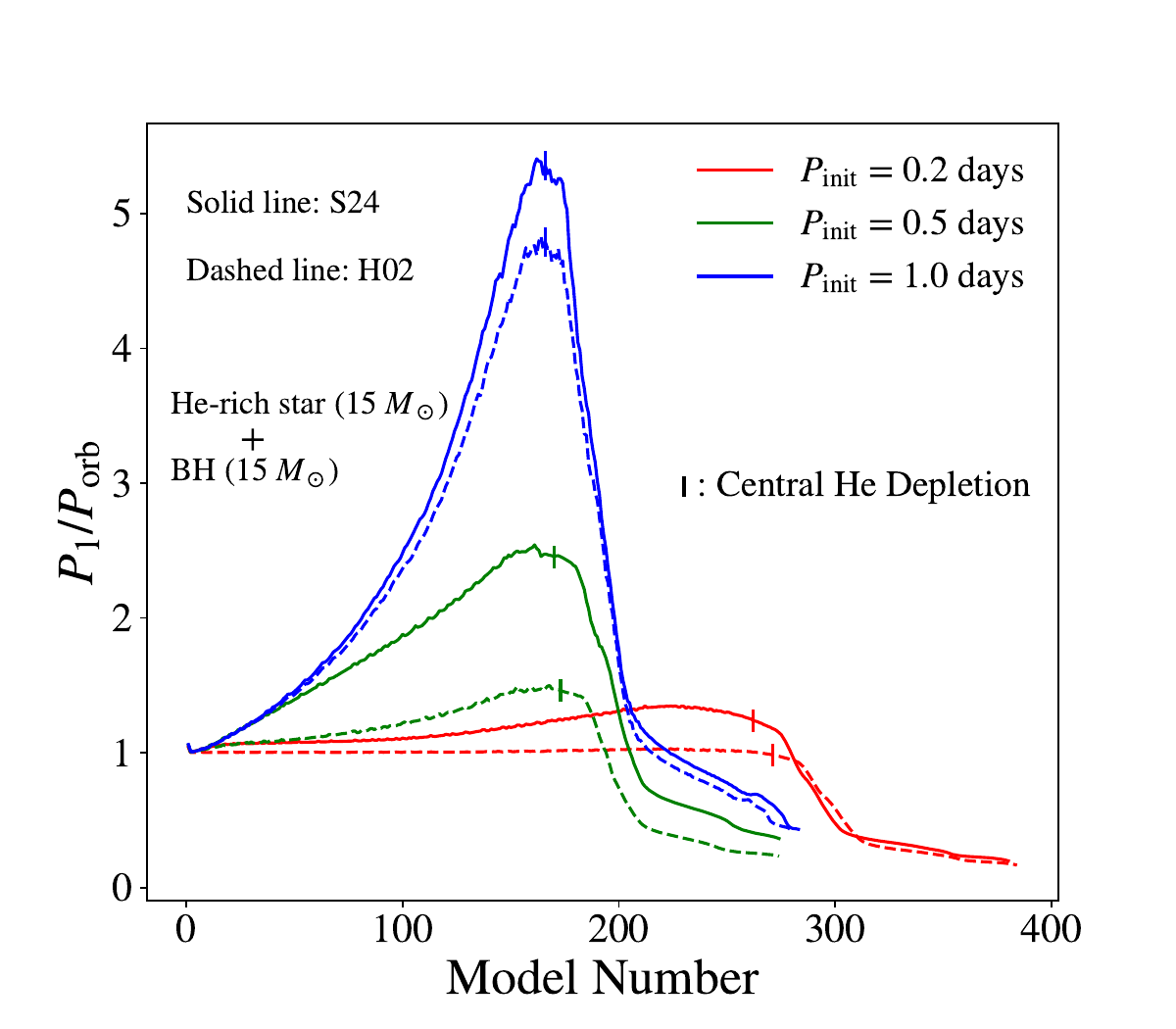}
     \includegraphics[width=0.99\columnwidth]{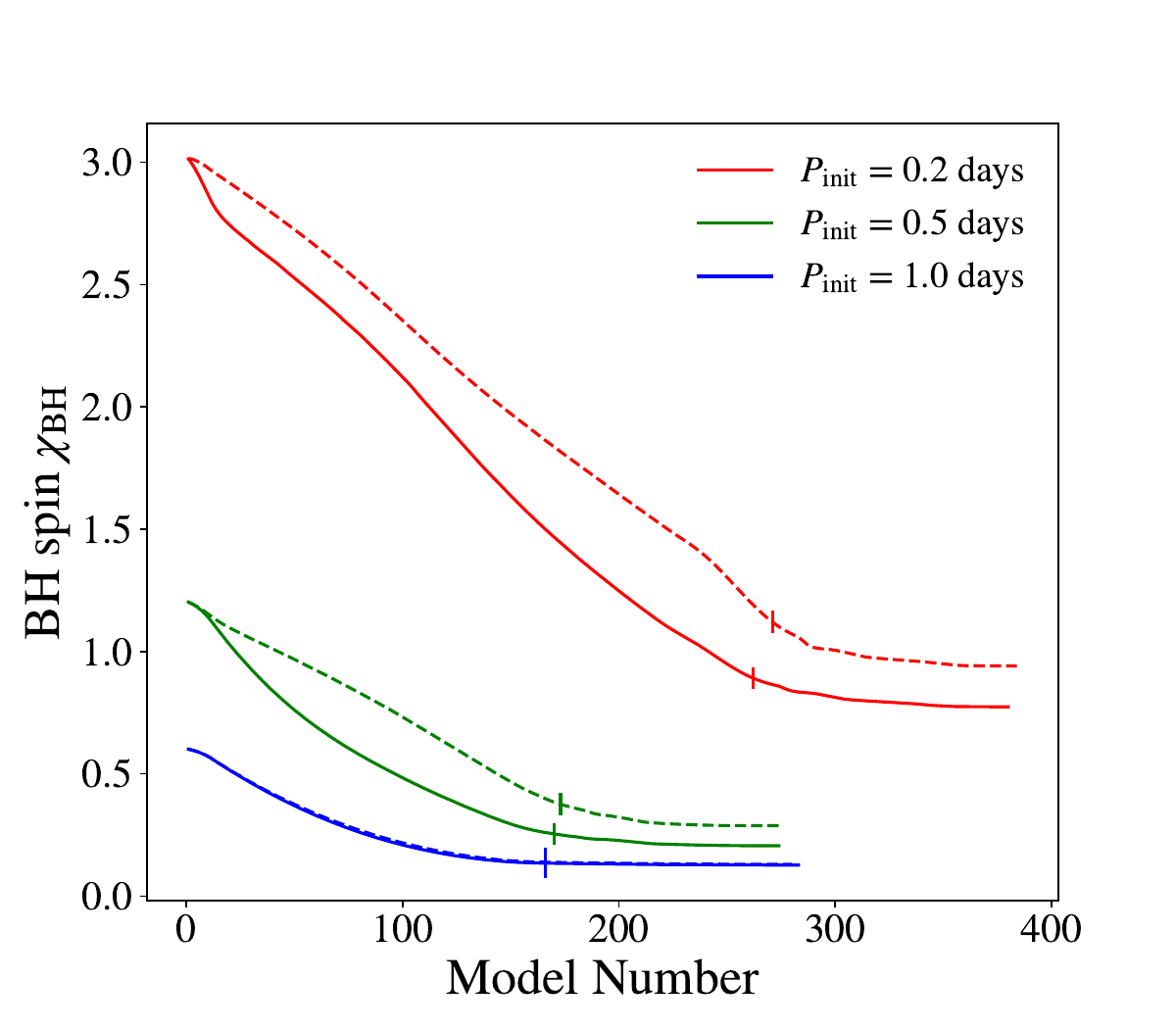}
     \caption{The ratio of the He-rich star's rotational period to the orbital period ($P_1/P_{\rm orb}$, \emph{left panel}) and the resulting BH spin ($\chi_{\rm BH}$, \emph{right panel}) are plotted as functions of model number for different initial orbital periods (red: 0.2 days; green: 0.5 days; blue: 1.0 days). Two synchronization timescales are compared: the solid line represents $\tau_{\rm sync}$ from \cite{Sciarini2024}, while the dashed line corresponds to $\tau_{\rm sync}$ from \cite{Hurley2002}. Vertical lines indicate the point at which the He-rich star reaches central helium depletion.}
     \label{tides}
\end{figure*}

In Figure \ref{tides}, we distinguish the models computed with the newly proposed $\tau_{{\rm sync}}$ as S24 and those using the default $\tau_{{\rm sync}}$ as H02 (see Equation (8) in \cite{Sciarini2024}). The left panel illustrates the spin-up of the He-rich star following central helium depletion, with varying $\tau_{{\rm sync}}$ values. Comparatively, the tidal effects using the newly proposed $\tau_{{\rm sync}}$ \citep{Sciarini2024} are weaker than those with the default one \citep{Hurley2002}. The right panel shows the resulting BH spin, assuming the He-rich star undergoes direct collapse to form a BH. In scenarios with weaker tides, the BH spin magnitude is reduced in close binaries (e.g., with orbital periods shorter than 1.0 days).
\end{appendix}

\end{document}